\documentclass[preprint,twoside,web]{ieeecolor}
\usepackage{tmi}
\usepackage{cite}
\usepackage{amsmath,amssymb,amsfonts}
\usepackage{graphicx}
\usepackage{textcomp}
\usepackage{subfig}
\usepackage{dsfont}
\usepackage{color, soul}
\usepackage{hyperref}
\usepackage{multirow}
\usepackage{algorithmicx,algpseudocode,algorithm}
\usepackage{booktabs}
\usepackage{float}

\makeatletter
\usepackage{xspace}
\def\@onedot{\ifx\@let@token.\else.\null\fi\xspace}
\DeclareRobustCommand\onedot{\futurelet\@let@token\@onedot}
\newcommand{\eqnref}[1]{Eq\onedot~\eqref{#1}}
\newcommand{\figref}[1]{Fig\onedot~\ref{#1}}
\newcommand{\algoref}[1]{Alg\onedot~\ref{#1}}

\newtheorem{thm}{Theorem}[section]

\newtheorem{assump}{Assumption}

\def\BibTeX{{\rm B\kern-.05em{\sc i\kern-.025em b}\kern-.08em
    T\kern-.1667em\lower.7ex\hbox{E}\kern-.125emX}}
\begin{document}
\title{High-Frequency Space Diffusion Model for Accelerated MRI}
\author{Chentao Cao, Zhuo-Xu Cui, Yue Wang, Shaonan Liu, Taijin Chen, Hairong Zheng, \IEEEmembership{Senior Member, IEEE},\\Dong Liang, \IEEEmembership{Senior Member, IEEE}, Yanjie Zhu, \IEEEmembership{Member, IEEE}
\thanks{Chentao Cao, Zhuo-Xu Cui and Yue Wang contributed equally to this manuscript.}
\thanks{Corresponding author: yj.zhu@siat.ac.cn (Yanjie Zhu).}
\thanks{Chentao Cao, Taijin Chen, Hairong Zheng, Dong Liang and Yanjie Zhu are with Lauterbur Research Center for Biomedical Imaging, Shenzhen Institute of Advanced Technology, Chinese Academy of Sciences, Shenzhen, China.}
\thanks{Chentao Cao and Taijin Chen are also with the School of Artificial Intelligence, the University of Chinese Academy of Sciences, Beijing, China.}
\thanks{Zhuo-Xu Cui and Dong Liang are with Research Center for Medical AI, Shenzhen Institutes of Advanced Technology, Chinese Academy of Sciences, Shenzhen, Chin.}
\thanks{Yanjie Zhu and Dong Liang are also with the National Center for Applied Mathematics
Shenzhen (NCAMS), Shenzhen 518000, China.}
\thanks{Dong Liang is also with Pazhou Lab, Guangzhou, China.}
\thanks{Yue Wang is with the School of Biomedical Engineering, Shenzhen University Medical School, Shenzhen University,  Shenzhen, China.}
\thanks{Shaonan Liu is with Computer Vision Institute, College of Computer Science and Software Engineering, Shenzhen University, Shenzhen, China. }
}
\maketitle

\begin{abstract}
Diffusion models with continuous stochastic differential equations (SDEs) have shown superior performances in image generation. It can serve as a deep generative prior to solving the inverse problem in magnetic resonance (MR) reconstruction. However, low-frequency regions of $k$-space data are typically fully sampled in fast MR imaging, while existing diffusion models are performed throughout the entire image or $k$-space, inevitably introducing uncertainty in the reconstruction of low-frequency regions. Additionally, existing diffusion models often demand substantial iterations to converge, resulting in time-consuming reconstructions. To address these challenges, we propose a novel SDE tailored specifically for MR reconstruction with the diffusion process in high-frequency space (referred to as HFS-SDE). This approach ensures determinism in the fully sampled low-frequency regions and accelerates the sampling procedure of reverse diffusion. Experiments conducted on the publicly available fastMRI dataset demonstrate that the proposed HFS-SDE method outperforms traditional parallel imaging methods, supervised deep learning, and existing diffusion models in terms of reconstruction accuracy and stability. The fast convergence properties are also confirmed through theoretical and experimental validation. Our code and weights are available at \href{https://github.com/Aboriginer/HFS-SDE}{https://github.com/Aboriginer/HFS-SDE}.
\end{abstract}


\section{Introduction}
\IEEEPARstart{M}{agnetic} resonance imaging (MRI) is a powerful tool of medical imaging but suffers from a long acquisition time. The main strategy for accelerating MR scan is to undersample the $k$-space data and then reconstruct images based on priors of images and measurements, as in parallel imaging and compressed sensing (CS)\cite{haldar2010compressed,  jin2016general, majumdar2015improving, pruessmann1999sense, grappa, lustig2010spirit, kt-SLR, Low-Rank, zhao2012image, huang2023deep}. 

In recent years, deep learning (DL)-based reconstruction methods have grown in popularity and shown great potential to remove aliasing artifacts caused by undersampling\cite{huang2023vigu, huang2022unsupervised, Wang, liang2020deep, han2019k, peng2022deepsense, oh2020unpaired, nakarmi2020multi, huang2019deep}. One broad family of DL-based methods is the supervised learning technique. This technique takes measured $k$-space data and images to be reconstructed as the inputs and outputs of a deep neural network and directly learns the mapping between them. Typical network architectures are based on unrolling that unroll an iterative solution of CS optimization into a network by learning the hyper-parameters and regularizations, e.g., ISTA-Net, ADMM-Net\cite{modl, sun2016deep,zhang2018ista, cui2021equilibrated}. Supervised learning techniques are usually trained for a specific imaging model or anatomy and show excellent performance in these tasks. However, generalization is one of the major bottlenecks of supervised learning methods. The trained network may show severely degraded reconstruction quality when applied out of distribution.

Another approach of DL-based reconstruction methods is the unsupervised learning technique based on distribution learning. It uses the data distribution predicted by a deep generative model as priors to solve the inverse problem of MR reconstruction. This technique adapts to changes in measurements easily because they are trained without references to measurements. Recently, diffusion models have shown significant progress and have become the new state-of-the-art (SOTA) deep generative models\cite{dhariwal2021diffusion, nichol2021improved, rombach2022high, yang2022diffusion}. Two successful classes of diffusion models, \textit{i.e.}, denoising diffusion probabilistic models (DDPMs)\cite{DDPM} and score matching with Langevin dynamic (SMLD)\cite{score-based}, have gained wide interest and demonstrated remarkable synthesis quality. Song et al. \cite{score-based-SDE} encapsulated these two approaches into a generalized framework of score-based generative models based on continuous stochastic differential equations (SDEs), improving the capabilities of diffusion models. SMLD and DDPM correspond to Variance Exploding (VE) SDE and Variance Preserving (VP) SDE, respectively. Under the framework of score-based SDEs, data distribution is transformed to a tractable distribution (\textit{i.e.}, Gaussian noise distribution) by a forward SDE. This process can be reversed for sample generation based on a reverse-time SDE derived from the forward SDE, given the score function of the marginal probability densities. Score-based SDEs have been adopted to the inverse problem of MR reconstruction and have shown impressive results\cite{jalal2021robust, song2022solving, chung2022score, kim2022diffusion, cui2022self, dar2022adaptive}.

However, the existing VE- and VP-SDEs were not proposed specifically for MR reconstruction and may lead to some pitfalls in diffusion model-based MR reconstructions. Conventional VE- and VP-SDEs are not designed specifically for MR reconstruction. In contrast to natural image processing scenarios, MR imaging often acquires fully sampled low-frequency regions in $k$-space for calibration. Executing VP- and VE-SDE models involves the diffusion process across the entire image domain, which cannot guarantee consistency between the low-frequency regions derived from the reconstructed image and the acquired one, thereby increasing reconstruction uncertainty. Additionally, both VE- and VP-SDE-based diffusion models suffer from slow convergence, requiring thousands of iterations to produce high-quality samples. The resulting long reconstruction time hinders the clinical use of fast MRI imaging based on VP- and VE-SDE diffusion models.

Rethinking MR reconstructions, subspace modeling methods often exhibit more robust and accurate performance compared to full-space modeling methods \cite{nguyen2019frequency, blaimer2004smash,10244070}. 
What's more, it is worth noting that the choice of the initial value significantly influences the convergence rate of iterative optimization algorithms. Starting from an optimized subspace often exhibits faster convergence speeds than a random initialized one \cite{kressner2014low,9632354}. These findings inspire us that subspace-based diffusion models may be more suitable for MR reconstruction.

In view of the above considerations, this study introduces a new SDE that specifically targets the diffusion process in high-frequency subspace, referred to as High-Frequency Space Stochastic Differential Equation (HFS-SDE). The primary objective of this approach is to ensure consistency between the reconstructed image and the fully sampled low-frequency $k$-space data, ultimately enhancing the robustness and accuracy of the reconstruction. Furthermore, its reverse process initiates from a low-frequency subspace accompanied by high-frequency noise, resulting in fewer iterations to convergence compared to existing VP- and VE-SDEs, which initiate from Gaussian noise. Finally, we provide a theoretical discussion that the proposed HFS-SDE exhibits a smaller weak convergence upper bound than the SDE in the full space. Experimental validation in MR reconstruction confirms that HFS-SDE can achieve superior performance and faster convergence speeds.

\subsection{Contributions}
\begin{enumerate}
    \item  We propose a new SDE (HFS-SDE) for diffusion model-based MR reconstruction with high stability. When applied to multi-coil MR data, we achieve a 12-fold uniform acceleration in 2D imaging. 
    \item We have proved theoretically that the HFS-SDE exhibits a smaller weak convergence upper bound compared to the SDE in the full space, as well as validated the accelerated convergence effect of the HFS-SDE in experiments through high-quality reconstructions achieved with only 100 sampling steps.
    \item The experiments show that more image details are preserved using HFS-SDE than VE- and VP-SDEs since the diffusion process focuses on high-frequency. 
\end{enumerate}

The following sections of the paper are organized as follows: Section \ref{Background} introduces the background, Section \ref{theory and methods} describes the proposed method, and Section \ref{experiments} provides the experimental results. Discussion and conclusion are given in Section \ref{discussion} and Section \ref{conclusions}.

\section{Background}\label{Background}
\subsection{Problem Formulation}
The imaging model of MR reconstruction can be formulated as:
\begin{equation}
    \mathbf{y}=\mathbf{A}\mathbf{x}+\boldsymbol{\epsilon},
    \label{MR forward model}
\end{equation}
where $\mathbf{y}$ is the undersampled measurements in the frequency domain~(\textit{i.e.}, $k$-space), $\mathbf{x}$ is the image to be reconstructed, $\mathbf{A}$ denotes the encoding matrix, $\mathbf{A}=\mathbf{M}\cdot \mathbf{F} \cdot \text{csm}$, $\mathbf{M}$ is the undersampling operator, $\mathbf{F}$ denotes the Fourier operator, $\text{csm}$ denotes the coil sensitivity, and $\boldsymbol{\epsilon} \sim \mathcal{N}(0, \sigma^2_\epsilon)$. For 2D image, $\mathbf{x} \in \mathbb{C}^n$, $\mathbf{y} \in \mathbb{C}^m$ and $\mathbf{A} \in \mathbb{C}^{m\times n}$.

\eqnref{MR forward model} can be formulated as an optimization problem:
\begin{equation}
    \mathbf{x}^*=\underset{\mathbf{x}}{\arg \min} \frac{1}{2}\|\mathbf{Ax}-\mathbf{y}\|_2^2+\mathcal{R}(\mathbf{x}),
    \label{MR optimization problem}
\end{equation}
where $\mathcal{R}(\mathbf{x})$ is the prior constraint of the MR image.

\subsection{Score-based SDE}
In score-based SDE\cite{score-based-SDE},  the diffusion process $\{\mathbf{x}(t)\}_{t=0}^T$ is continuous with a time variable $t \in[0, T]$, and thus can be constructed as the solution of an SDE. The corresponding reverse process can be accurately modeled with a reverse-time SDE of the forward one, extending the capability of diffusion models. The reverse-time SDE can be approximated by training a network to estimate the score of the densities at each time step and then used for sample generation using numerical SDE solvers.  

The diffusion process can be described as the solution of the following SDE:
\begin{equation}
    \mathrm{d} \mathbf{x}=\mathbf{f}(\mathbf{x}, t) \mathrm{d} t+g(t) \mathrm{d} \mathbf{w},
\end{equation}
where function $\mathbf{f}$ is the drift function of $\mathbf{x}(t)$ and $g$ is called the diffusion coefficient. $\mathbf{w}$ is the standard Wiener process. The reverse process is also a diffusion process, modeled by the following reverse-time SDE:
\begin{equation}
    \mathrm{d} \mathbf{x}=\left[\mathbf{f}(\mathbf{x}, t)-g(t)^{2} \nabla_{\mathbf{x}} \log p_{t}(\mathbf{x})\right] \mathrm{d} t+g(t) \mathrm{d} \mathbf{\bar w},
    \label{related-work reverse sde}
\end{equation}
where $\nabla_{\mathbf{x}} \log p_{t}(\mathbf{x})$ is known as the score function, $\mathbf{\bar w}$ is the standard Wiener process when the time goes back to $0$ from $T$. 

In order to solve \eqnref{related-work reverse sde}, the score function $\nabla_{\mathbf{x}} \log p_{t}(\mathbf{x})$ can be estimated by a neural network $\mathbf{s}_{\boldsymbol{\theta}}(\mathbf{x}(t), t)$ with the loss function of 
\begin{multline}
        \boldsymbol{\theta}^{*}=\underset{\boldsymbol{\theta}}{\arg \min } \mathbb{E}_{t}\Big\{\lambda(t) \mathbb{E}_{\mathbf{x}(0)} \mathbb{E}_{\mathbf{x}(t) \mid \mathbf{x}(0)}\big[\big\|\mathbf{s}_{\boldsymbol{\theta}}(\mathbf{x}(t), t)\\-\nabla_{\mathbf{x}(t)} \log p_{0 t}(\mathbf{x}(t) \mid \mathbf{x}(0))\big\|_{2}^{2}\big]\Big\},
    \label{SDE-loss}
\end{multline}
where $\lambda(t)$ is a positive weighting function, $t$ is uniformly sampled over $[0, T]$, $\mathbf{x}(0) \sim p_{\text{data}}$ and $\mathbf{x}(T) \sim p_{\text{noise}}$. $p_{0 t}(\mathbf{x}(t) \mid \mathbf{x}(0))$ is the perturbation kernel. Once the score model $\mathbf{s}_{\boldsymbol{\theta}}(\mathbf{x}(t), t)$ is trained by \eqnref{SDE-loss}, the image can be generated by the reverse diffusion process.

The two most commonly used SDEs are VE-SDE and VP-SDE below in diffusion models.
\subsubsection{VE-SDE}
The forward process of VE-SDE is: 
\begin{equation}
    \mathrm{d} \mathbf{x}=\sqrt{\frac{\mathrm{d}\left[\sigma^2(t)\right]}{\mathrm{d} t}} \mathrm{~d} \mathbf{w}.
\end{equation}
During the forward diffusion process, VE-SDE adds Gaussian noise with zero mean and gradually increasing variance to the data $\mathbf{x}(0)$. The perturbation kernel $p_{0 t}(\mathbf{x}(t) \mid \mathbf{x}(0))$ of VE-SDE is:
\begin{equation}
p_{0 t}(\mathbf{x}(t) \mid \mathbf{x}(0))=\\ 
    \mathcal{N}(\mathbf{x}(t) ; \mathbf{x}(0),(\sigma^2(t)-\sigma^2(0)) \mathbf{I}),
\label{VE perturbation kernel}    
\end{equation}
where $\mathbf{x}(t)$ denotes the perturbed data at the moment $t$, $\mathbf{x}(0) \sim p_0(\mathbf{x})$. The perturbation kernel indicates the mean of $\mathbf{x}(t)$ is always the same as $\mathbf{x}(0)$. $\sigma(t)$ is the given coefficient to control the noise level.
\subsubsection{VP-SDE}
The forward process of VP-SDE is: 
\begin{equation}
    \mathrm{d} \mathbf{x}=-\frac{1}{2} \beta(t) \mathbf{x} \mathrm{d} t+\sqrt{\beta(t)} \mathrm{d} \mathbf{w}.
\end{equation}
Unlike VE-SDE, VP-SDE gradually attenuates the signal during the forward diffusion process until the mean value reaches 0. The perturbation kernel $p_{0 t}(\mathbf{x}(t) \mid \mathbf{x}(0))$ of VP-SDE is:
\begin{multline}
    p_{0 t}(\mathbf{x}(t)\mid \mathbf{x}(0))
    = \mathcal{N}\big(\mathbf{x}(t) ; \mathbf{x}(0) e^{-\frac{1}{2} \int_0^t \beta(s) \mathrm{d} s},\\ \mathbf{I}-\mathbf{I} e^{-\int_0^t \beta(s) \mathrm{d} s}\big).
    \label{VP perturbation kernel} 
\end{multline}
$\beta(s)$ is the given coefficient to control the noise level.

\subsection{Score-based SDEs for MRI Reconstruction}\label{Score-based SDEs for MRI Reconstruction}
Score-based SDEs have been applied for MR reconstruction\cite{chung2022score, song2022solving, chung2022come, tu2022wkgm, peng2022one, huang2023cdiffmr}. Specifically, the stochastic samples from the image distribution can be generated via posterior sampling instead of directly modeling the prior information as in \eqnref{MR optimization problem}.

Given the undersampled $k$-space data $\mathbf{y}$, MR images can be generated through the reverse-time SDE:
\begin{equation}
\mathrm{d} \mathbf{x}=\left\{\mathbf{f}(\mathbf{x}, t)-g(t)^2\nabla_{\mathbf{x}} \log p_t(\mathbf{x} \mid \mathbf{y})\right\} \mathrm{d} t+g(t) \mathrm{d} \overline{\mathbf{w}},
\end{equation}
where $\mathbf{x}$ is the image to be reconstructed.

Taking VP-SDE as an example, the iterative rule of the reverse-time SDE is
\begin{multline}
        \mathbf{x}_{i}=\mathbf{x}_{i+1}+\frac{1}{2} \beta_{i+1} \mathbf{x}_{i+1}+\beta_{i+1}         \mathbf{s}_{\theta^{*}}\left(\mathbf{x}_{i+1}, i+1\right)+\sqrt{\beta_{i+1}} \mathbf{z}_{i}, \\\quad i=0,1, \cdots, N-1,
        \label{VP-SDE-d}
\end{multline}
where $\mathbf{x}_0$ is considered to be the reconstructed MR image, $\mathbf{s}_{\theta^{*}}\left(\mathbf{x}_{i+1}, i+1\right)$ is the trained score-based model, $\mathbf{z}_{i} \sim \mathcal{N}(\mathbf{0}, \mathbf{I})$ and $\{\beta_{1}, \beta_{2}, \cdots, \beta_{N}\}$ are given coefficients to control the noise level. 
Executing SDE (\eqnref{VP-SDE-d}) involves diffusion across the entire image domain, which cannot guarantee consistency between the low-frequency regions of the reconstructed image and the deterministic fully sampled low-frequency regions, thereby increasing imaging uncertainty. Additionally, the iterates (\eqnref{VP-SDE-d}) often require thousands of iterations to converge, resulting in long computational times.

It is worth noting that we have mentioned two types of sampling above. One is undersampling in $k$-space to reduce the acquisition time of MRI, indicated by the undersampling operator $\mathbf{M}$. The other is posterior sampling, which refers to sampling from the distribution to obtain reconstructed MR images during the reverse diffusion process.

\section{Theory and methods}\label{theory and methods}
This section introduces the proposed HFS-SDE first, then describes how to train the score network and reconstruct MR images according to the reverse-time HFS-SDE. Fig. \ref{fig:general} shows the framework of the HFS-SDE-based MR reconstruction. 
\subsection{The High-Frequency Space SDE}
\begin{figure*}[!t]
    \centerline{\includegraphics[width=\textwidth]{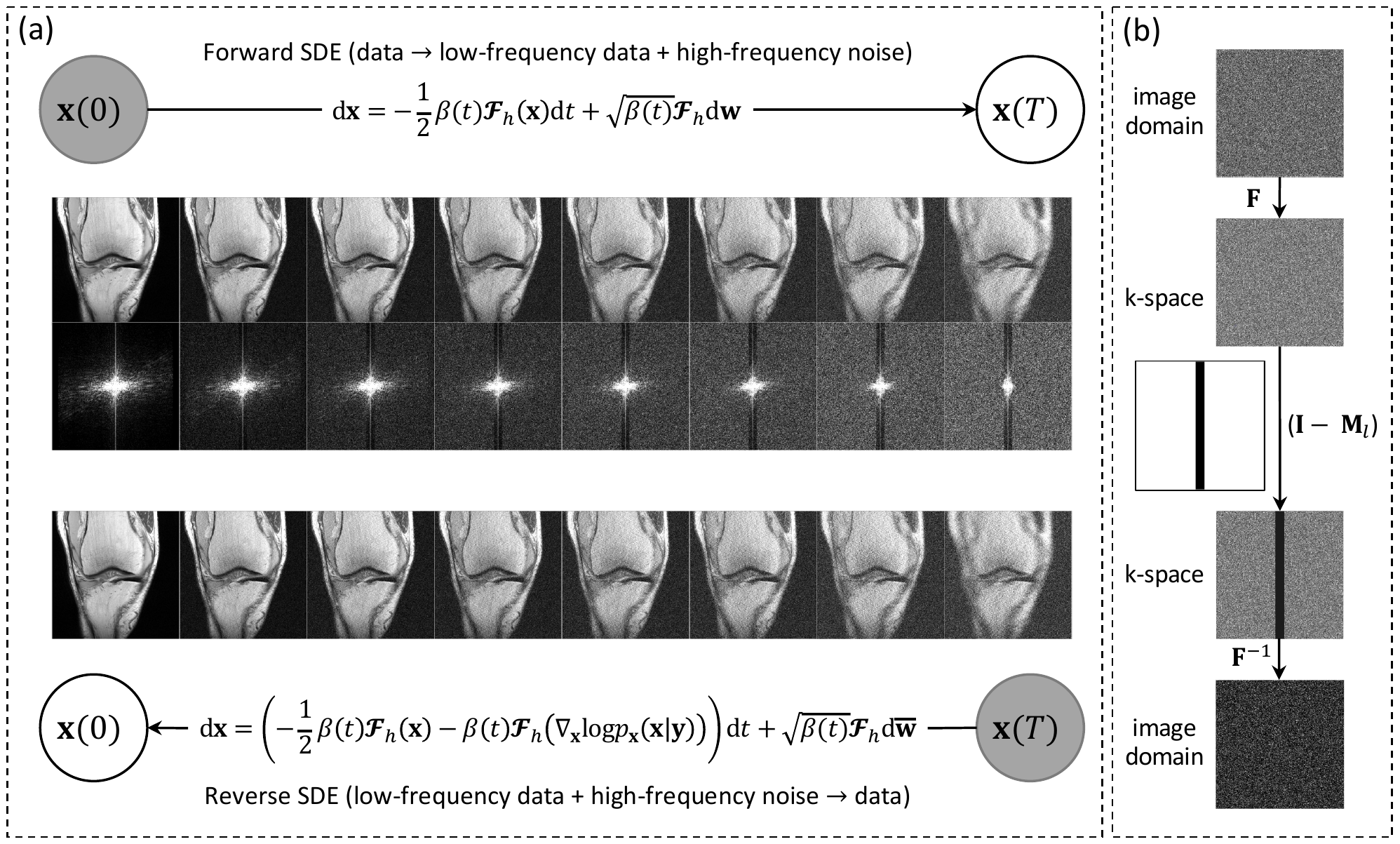}}
    \caption{Illustrate the proposed high-frequency space diffusion model. (a) In the forward process, high-frequency noises with multiple scales are added to the data (first row). The second row shows the $k$-space corresponding to the perturbed data (no noise is added to the central region of $k$-space). 
    (b) Demonstrate the steps to acquire high-frequency noise. High-frequency noise is added to the data in the image domain.}
    \label{fig:general}
\end{figure*}

\begin{figure}[!t]
    \centering
    \resizebox{0.95\columnwidth}{!}{\includegraphics{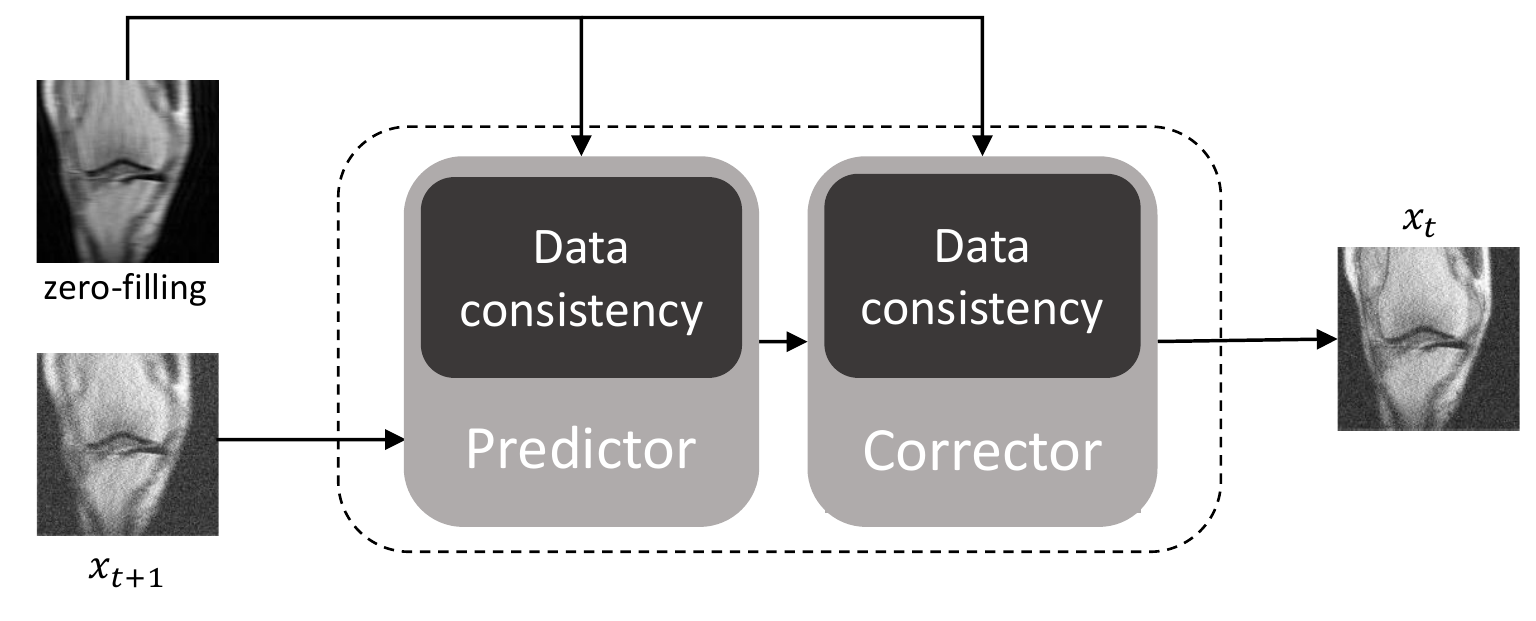}}
    \caption{Predictor corrector sampling. The predictor includes the original predictor\cite{score-based-SDE} and data consistency priors, while the corrector comprises the original corrector\cite{score-based-SDE} and data consistency priors. Zero-filling represents undersampled images with $k$-space zero-filling.}
    \label{fig:pcsample}
\end{figure}

To restrict the diffusion process in high-frequency space,  we first construct high- and low-frequency operators $\boldsymbol{\mathcal{F}}_h(\cdot)$ and $\boldsymbol{\mathcal{F}}_l(\cdot)$ as follows:
\begin{equation}
    \left\{\begin{array}{l}
    \boldsymbol{\mathcal{F}}_h(\cdot):=\mathbf{F}^{-1}((\mathbf{I} - \mathbf{M}_l)\cdot \mathbf{F}(\cdot))\\
    \boldsymbol{\mathcal{F}}_l(\cdot):=\mathbf{F}^{-1}(\mathbf{M}_l \cdot \mathbf{F}(\cdot))
    \end{array}\right.,
\end{equation}
where $\mathbf{M}_l$ is the mask to differentiate low- and high-frequency spaces, with value $1$ for $k$-space center (
1D mask with a size of $n_l$) and $0$ for the rest of the region (high-frequency region). $\mathbf{I}$ denotes all-ones matrix ($(\mathbf{I}-\mathbf{M}_l$) is shown in Fig. \ref{fig:general} (b)). In an extreme case, $n_l=0$, HFS-SDE is downgraded to VP-SDE, and in this study, we choose $n_l=16$. In the multi-coil scenario, 
\begin{equation}
    \left\{\begin{array}{l}
    \boldsymbol{\mathcal{F}}_h(\mathbf{x}):=\sum_{j=1}^{n} \text{csm}_j^{*} \cdot \mathbf{F}^{-1}\left((\mathbf{I}-\mathbf{M}_l)\cdot \mathbf{F}(\text{csm}_j\cdot \mathbf{x}_i)\right)\\
    \boldsymbol{\mathcal{F}}_l(\mathbf{x}):=\sum_{j=1}^{n} \text{csm}_j^{*} \cdot \mathbf{F}^{-1}\left(\mathbf{M}_l \cdot \mathbf{F}(\text{csm}_j\cdot \mathbf{x}_i)\right)
    \end{array}\right.,
\end{equation}
where $n$ indicates the number of coils, $\text{csm}_j$ denotes the sensitivity of the $j$th coil and $\mathbf{x}$ denotes the MR image. $\boldsymbol{\mathcal{F}}_h$ is the operator that extracts the high-frequency component of $\mathbf{x}$. Specifically, it first transforms the MR image $\mathbf{x}$ to the frequency domain, followed by applying a high-frequency mask~($\mathbf{I}-\mathbf{M}_l$), and subsequently transforms back to the image domain. $\boldsymbol{\mathcal{F}}_l(\mathbf{x})$ signifies a similar process as $\boldsymbol{\mathcal{F}}_h(\mathbf{x})$ except for the low-frequency mask $\mathbf{M}_l$ is applied in the frequency domain. 

Assuming that there are $N$ noise scales in the forward diffusion process constructed by the following discrete Markov chain,
\begin{multline}
    \mathbf{x}_{i}=\boldsymbol{\mathcal{F}}_l(\mathbf{x}_{i-1}) + \sqrt{1-\beta_{i}} \boldsymbol{\mathcal{F}}_h(\mathbf{x}_{i-1})+\sqrt{\beta_{i}} \boldsymbol{\mathcal{F}}_h(\mathbf{z}_{i-1}), \\ \quad i=1, \cdots, N,
    \label{HFS discrete Markov}
\end{multline}
where $\{\beta_{1}, \beta_{2}, \cdots, \beta_{N}\}$ are given coefficients to control the noise level and $\mathbf{z}_{i-1} \sim \mathcal{N}(\mathbf{0}, \mathbf{I})$. As $\Delta t=\frac{1}{N} \rightarrow 0$, \eqnref{HFS discrete Markov} converges to the following SDE (detailed in Appendix \ref{appendix:A}):
\begin{equation}
    \mathrm{d} \mathbf{x}=-\frac{1}{2} \beta(t) \boldsymbol{\mathcal{F}}_h(\mathbf{x}) \mathrm{d} t+\sqrt{\beta(t)} \boldsymbol{\mathcal{F}}_h\mathrm{d}\mathbf{w}.
    \label{forward sde}
\end{equation}

By S{\"a}rkk{\"a}, Simo \cite{sarkka2019applied} and let $\beta(t)=\bar{\beta}_{\min }+t\left(\bar{\beta}_{\max }-\bar{\beta}_{\min }\right)$ for $t \in [0, 1]$, the perturbation kernel of  HFS-SDE can be derived as
\begin{multline}
    \begin{aligned}
    &~~~p_{0 t}(\mathbf{x}(t)\mid \mathbf{x}(0)) \\
    &= \mathcal{N}\big(\mathbf{x}(t) ; e^{(-\frac{1}{4} t^{2}\left(\bar{\beta}_{\max }-\bar{\beta}_{\min }\right)-\frac{1}{2} t \bar{\beta}_{\min})\boldsymbol{\mathcal{F}}_h} \mathbf{x}(0), \\&~~~~~~~~~~~~~~\mathbf{I}-\mathbf{I} e^{(-\frac{1}{2} t^{2}\left(\bar{\beta}_{\max }-\bar{\beta}_{\min }\right)-t \bar{\beta}_{\min })\boldsymbol{\mathcal{F}}_h}\big), \quad t \in[0,1],
    \end{aligned}
\end{multline}
where the mean $\boldsymbol{\mu}$ and covariance matrix $\boldsymbol{\Sigma}$ are exponential functions with the high-frequency operator $\boldsymbol{\mathcal{F}}_h$ in the exponent. 
The calculation of the exponential operator requires a spectral decomposition of $\boldsymbol{\mathcal{F}}_h$. Due to the large dimension of $\boldsymbol{\mathcal{F}}_h$, the computational complexity of the spectral decomposition is intolerable. Therefore, we proposed an approximate method to calculate it. Let the coefficient before $\boldsymbol{\mathcal{F}}_h$ to be $k$. The exponential function with $\boldsymbol{\mathcal{F}}_h$ can be expanded using Taylor expansion:
\begin{equation}
    \begin{aligned}
        e^{k\boldsymbol{\mathcal{F}}_h} &= \prod \limits_{i=0}^L e^{\frac{k}{L}\boldsymbol{\mathcal{F}}_h} = \prod \limits_{i=0}^L\left(\mathbf{I}+\frac{k}{L}\boldsymbol{\mathcal{F}}_h + o(\frac{1}{L})\boldsymbol{\mathcal{F}}_h^2 \right).
    \end{aligned}
\end{equation}
This function involves massive calculations and is still difficult to implement. Thanks to $\boldsymbol{\mathcal{F}}_h(\cdot)^2=\boldsymbol{\mathcal{F}}_h(\cdot)$, we can get
\begin{equation}
    \begin{aligned}
        \lim\limits_{L\rightarrow\infty}\prod \limits_{i=0}^L\left(\mathbf{I}+\frac{k}{L}\boldsymbol{\mathcal{F}}_h + o(\frac{1}{L})\boldsymbol{\mathcal{F}}_h^2 \right)=(e^k-1)\boldsymbol{\mathcal{F}}_h + \mathbf{I}.
    \end{aligned}
\end{equation}
Then the mean $\boldsymbol{\mu}$ and covariance matrix $\boldsymbol{\Sigma}$ of the perturbation kernel can be re-expressed as
\begin{equation}
    \left\{\begin{array}{l}
    \boldsymbol{\mu}=\left(e^{(-\frac{1}{4} t^{2}\left(\bar{\beta}_{\max }-\bar{\beta}_{\min }\right)-\frac{1}{2} t \bar{\beta}_{\min})}-1\right)\boldsymbol{\mathcal{F}}_h(\mathbf{x}(0))+\mathbf{x}(0)\\
    \boldsymbol{\Sigma}=\left(1-e^{-\frac{1}{2} t^{2}\left(\bar{\beta}_{\max }-\bar{\beta}_{\min }\right)-t \bar{\beta}_{\min }}\right)\boldsymbol{\mathcal{F}}_h
    \end{array}\right..
    \label{appendix mean and covariance}
\end{equation}

\subsection{Estimate Score for HFS-SDE} 
Once the perturbation kernel is determined, the score model $\mathbf{s}_{\boldsymbol{\theta}}(\mathbf{x}(t), t)$ can be trained by a network. Unlike VE- and VP-SDEs, the network no longer estimates Gaussian noise of the whole image but only predicts the high-frequency part of Gaussian noise (details in Appendix \ref{appendix:B}):
\begin{multline}
    \boldsymbol{\theta}^{*}=\underset{\boldsymbol{\theta}}{\arg \min } \mathbb{E}_{t}\big\{\lambda(t) \mathbb{E}_{\mathbf{x}(0)} \mathbb{E}_{\mathbf{x}(t) \mid \mathbf{x}(0)}\big[\big\|\boldsymbol{\mathcal{F}}_h(\mathbf{z})+\\\sqrt{\big(1-e^{-\frac{1}{2} t^{2}\left(\bar{\beta}_{\max }-\bar{\beta}_{\min }\right)-t \bar{\beta}_{\min }}\big)}\boldsymbol{\mathcal{F}}_h(\mathbf{s}_{\boldsymbol{\theta}}(\mathbf{x}(t), t))\big\|_{2}^{2}\big]\big\},
\end{multline}
where $\lambda(t)$ is the positive weighting function and $\mathbf{z}$ is Gaussian noise. Empirically, our network only predicts high-frequency noise, releasing the ability of the network.

\subsection{Reverse HFS-SDE for MR Reconstruction} 
Based on \eqnref{forward sde}, the conditional reverse-time HFS-SDE can be deduced as\cite{anderson1982reverse}:
\begin{multline}
    \mathrm{d} \mathbf{x}=\left(-\dfrac{1}{2}\beta(t) \boldsymbol{\mathcal{F}}_h(\mathbf{x})-\beta(t)\boldsymbol{\mathcal{F}}_h(\nabla_{\mathbf{x}} \log p_{t}(\mathbf{x} \mid \mathbf{y}))\right) \mathrm{d} t\\+\sqrt{\beta(t)}\boldsymbol{\mathcal{F}}_h\mathrm{d} \overline{\mathbf{w}}.
    \label{reverse sde}
\end{multline}
According to Bayes' theorem and \eqnref{MR forward model}, $\nabla_{\mathbf{x}} \log p_{t}(\mathbf{x}(t) \mid \mathbf{y})$ can be written as:
\begin{equation}
    \begin{aligned}
        \nabla_{\mathbf{x}} \log p_{t}(\mathbf{x}(t) \mid \mathbf{y}) & = \nabla_{\mathbf{x}} \log p_{t}(\mathbf{x}(t)) + \nabla_{\mathbf{x}} \log p_{t}(\mathbf{y} \mid \mathbf{x}(t))\\
        & \approx \mathbf{s}_{{\boldsymbol{\theta}}^*}(\mathbf{x}(t), t) + \nabla_{\mathbf{x}} \log p_{t}(\mathbf{y} \mid \mathbf{x}(t))\\
        & \approx \mathbf{s}_{{\boldsymbol{\theta}}^*}(\mathbf{x}(t), t) + \frac{\mathbf{A}^H(\mathbf{y}-\mathbf{A} \mathbf{x}(t))}{\sigma^2_\epsilon}.
    \end{aligned}
\end{equation}
where $\sigma^2_\epsilon$ is a hyperparameter that requires annealing; the specific settings can be found in our code. HFS-SDE generates samples through the predictor-corrector method from \cite{score-based-SDE}. The framework is illustrated in Fig. \ref{fig:pcsample}, and the specific algorithm is presented in \algoref{alg:PC Sampling-HFS}. Note that in each step of Predictor-Corrector Sampling, $\mathbf{x}_i$ is generated conditioned on $\mathbf{y}$. Thus, the final reconstruction will align with $\mathbf{y}$.

Compared to the Gaussian noise initialization of VP- or VE-SDE, HFS-SDE model initializes $\mathbf{x}_{T}$ using the low-frequency data $\mathbf{M}_l\mathbf{y}$. Empirically, initializing closer to the image to be reconstructed often accelerates the convergence of the iteration sequence. Next, we will theoretically explain this claim. 

We study how the expected value of any suitably smooth statistic of $\mathbf{x}(t)$ in HFS-SDE (\eqnref{reverse sde}) evolves in time. We can define the following generator: 
\begin{equation}
\mathcal{L}f(\mathbf{x}(t)):=\lim_{h\rightarrow0^+}\frac{\mathbb{E}[f(\mathbf{x}(t+h))]-f(\mathbf{x}(t))}{h}.
\end{equation}
Then, we have
\begin{equation}
\mathbb{E}[f(\mathbf{x}(0))]=e^{-h\mathcal{L} } \circ\cdots\circ e^{-h\mathcal{L} }f(\mathbf{x}(T)),
\end{equation}
where the operator $e^{T\mathcal{L} }$ is called the Kolmogorov operator for HFS-SDE (\eqnref{reverse sde}).  Given HFS-SDE (\eqnref{reverse sde}) with an invariant measure $p_0(\mathbf{x})$, the posterior average is defined as $\overline{\phi}:=\int\phi(\mathbf{x})p_0(\mathbf{x})$. In practice, the discrete form of HFS-SDE (\eqnref{reverse sde}) is illustrated in the ninth line of \algoref{alg:PC Sampling-HFS}. For the discrete samples $(\mathbf{x}_i:=\mathbf{x}(hi))_{i=0}^{N}$, we define a functional $\psi$ that solves the following Poisson equation:
\begin{equation}
\mathcal{L}\psi(\mathbf{x}_i)=\phi(\mathbf{x}_i)- \overline{\phi}
\end{equation}
for some test function $\phi$.
\begin{assump}\label{ass:1}
    $\psi$ and its up to 3rd-order derivatives, $\mathcal{D}^k \psi$, are bounded by a function $ \mathcal{V}$, i.e., $\left\|\mathcal{D}^k \psi\right\| \leq C_k \mathcal{V}^{p_k}$ for $k=(0,1,2,3), C_k, p_k>0$. Furthermore, the expectation of $\mathcal{V}$ on $\left\{\mathbf{x}_i\right\}$ is bounded: $\sup _l \mathbb{E} \mathcal{V}^p\left(\mathbf{x}_i\right)<\infty$, and $\mathcal{V}$ is smooth such that $\sup _{s \in(0,1)} \mathcal{V}^p(s \mathbf{x}+(1-s) \mathbf{z}) \leq$ $C\left(\mathcal{V}^p(\mathbf{x})+\mathcal{V}^p(\mathbf{z})\right), \forall \mathbf{x}, \mathbf{z}, p \leq \max \left\{2 p_k\right\}$ for some $C>0$.
\end{assump} 

\begin{thm}\label{thm:1}
 Suppose Assumption \ref{ass:1} holds. The discrete HFS-SDE exhibits a smaller weak convergence upper bound compared to the SDE in the full space, which is achieved by replacing the operator $\boldsymbol{\mathcal{F}}_h$ with the identity operator $\boldsymbol{\mathcal{I}}$.  
\end{thm}
The proof is presented in the Appendix.~\ref{app: Proof}. Thus, we have theoretically demonstrated the acceleration effect of modeling the diffusion model in high-frequency space.

\begin{algorithm}
	\caption{PC Sampling (HFS-SDE).}
	\label{alg:PC Sampling-HFS}
	\begin{algorithmic}[1]
	    \Require{$\{\beta_i\}_{i=1}^N, \{\alpha_i\}_{i=1}^N, \text{csm}, \mathbf{\hat{y}}, \lambda_1, \lambda_2, r, N, M, \mathbf{M_u}$.} \Comment{$\text{csm} = \{\text{csm}_1, \cdots, \text{csm}_n\}$, $M_u$ is the undersampling mask}
	   \State{$\mathbf{x}_{N} \sim \mathcal{N}(\mathbf{F}^{-1}(\mathbf{M}_l\mathbf{y}), \boldsymbol{\mathcal{F}}_h)$}
	    \For{$i = N-1$ to $0$}
	        \State{$\mathbf{z} \sim \mathcal{N}(\mathbf{0}, \mathbf{I})$}
	        \State{$\mathbf{g} \leftarrow \boldsymbol{\mathcal{F}}_h(\mathbf{s}_{\boldsymbol{\theta^*}}\left(\mathbf{x}_{i+1}, i+1\right)$})
	        \State{$\mathbf{G}=\sum_{j=1}^{n} \text{csm}_j^{*} \cdot \mathbf{F}^{-1}\left(\mathbf{F}(\text{csm} \cdot \mathbf{x}_i) \cdot \mathbf{M_u} - \mathbf{\hat{y}}\right)$}
	        \State{$\epsilon \leftarrow \lambda_1\left(\|\mathbf{g}\|_{2} /\|\mathbf{G}\|_{2}\right)$}
	        \State{$\mathbf{x}_{i} \leftarrow \mathbf{x}_{i+1}+\dfrac{1}{2}\beta_{i+1}\boldsymbol{\mathcal{F}}_h(\mathbf{x}_i)+\beta_{i+1}(\mathbf{g}-\epsilon\mathbf{G})+\sqrt{\beta_{i+1}}\boldsymbol{\mathcal{F}}_h(\mathbf{z})$}
            \For{$k \gets 1$ to $M$}
                \State{$\mathbf{z} \sim \mathcal{N}(\mathbf{0}, \mathbf{I})$}
                \State{$\mathbf{g} \leftarrow \boldsymbol{\mathcal{F}}_h(\mathbf{s}_{\boldsymbol{\theta}^*}\left(\mathbf{x}_{i}^{k-1}, i\right)$}) 
                \State{$\mathbf{G}=\sum_{j=1}^{n} \text{csm}_j^{*} \cdot \mathbf{F}^{-1}\left(\mathbf{F}(\text{csm} \cdot \mathbf{x}_i^k) \cdot \mathbf{M_u} - \mathbf{\hat{y}}\right)$}
                \State{$\epsilon_1 \leftarrow 2 \alpha_{i}\left(r\|\mathbf{z}\|_{2} /\|\mathbf{g}\|_{2}\right)^{2}$}
                \State{$\epsilon_2 \leftarrow \|\mathbf{g}\|_{2} /(\lambda_2 \cdot \|\mathbf{G}\|_{2})$}
                \State{$\mathbf{x}_{i}^{k} \leftarrow \mathbf{x}_{i}^{k-1}+\epsilon_1 (\mathbf{g}-\epsilon_2\mathbf{G})+\sqrt{2 \epsilon_1} \boldsymbol{\mathcal{F}}_h(\mathbf{z})$}
            \EndFor
            \State{$\mathbf{x}_{i-1}^{0} \leftarrow \mathbf{x}_{i}^{M}$}
        \EndFor
        \item[]
        \Return{${\mathbf{x}}_0^0$} \Comment{$\mathbf{x}_0 = \mathbf{x}_i^0$}
	\end{algorithmic}
\end{algorithm}

\section{Experiments}\label{experiments}
\subsection{Experimental Setup}
\subsubsection{Experimental Data}
We conducted experiments on the public fastMRI dataset\footnote{\url{https://fastmri.org/}}\cite{zbontar2018fastmri, knoll2020fastmri}. For the training dataset, we selected 973 individuals from the fastMRI multi-coil knee dataset and dropped the first six slices from each individual due to poor image quality. A total of 28904 T1- and T2-weighted images were finally involved in the training dataset. We selected two datasets for testing. One was the knee data from 3 individuals selected randomly from the validation dataset, and the other one was the brain data from 1 individual. The first six slices of each individual in the test data were also discarded. We finally got 95 T1-weighted knee images and 10 T2-weighted brain images. We cropped the image size to 320$\times$320, and normalized the data by its own standard deviation before inputting it into the score network. We estimated the sensitivity maps using ESPIRiT \cite{uecker2014espirit} implemented in the BART toolbox \cite{uecker2016bart} with a $24\times24$ $k$-space center region.

\subsubsection{Parameter Configuration}
We compared HFS-SDE with iterative SENSE\cite{pruessmann1999sense}, supervised deep learning method~(ISTA-Net\cite{zhang2018ista}, DeepCascade\cite{schlemper2017deep}, VarNet\cite{sriram2020end}), GAN-based method~(CycleGAN)\cite{oh2020unpaired}, and score-based model (VE-SDE and VP-SDE)\cite{score-based-SDE}, with the results of each method tuned to the best. VP-SDE can be regarded as the ablation method of HFS-SDE. In the training phase, ISTA-Net and DeepCascade models were trained for 50 epochs, while CycleGAN was trained for 100 epochs. VE-, VP-, and HFS-SDE models were all trained for 190 epochs. For these SDE-based methods, the noise scale was set to $N=1000$. The exponential moving average (EMA) rate was 0.9999 for VP- and HFS-SDE, and 0.999 for VE-SDE. VP-SDE and HFS-SDE controlled the noise level during forward diffusion by setting $\beta_{max}=20$, and $\beta_{min}=0.1$ in \algoref{alg:PC Sampling-HFS}, while VE-SDE controlled the noise level in forward diffusion by setting $\sigma_{max}=348$ and $\sigma_{min}=0.1$\cite{song2020improved}. 
The network used in  VE-, VP-, and HFS-SDEs was the U-net architecture improved by Song et al.(\textit{i.e.}, \texttt{ddpm}\footnote{\url{https://github.com/yang-song/score_sde_pytorch}\label{sde code}} in the code of score-based SDEs). Since MRI data is complex-valued, it is split into real and imaginary components before being fed into the network, with each component serving as a separate channel for dual-channel real data input.

\subsubsection{Performance Evaluation}
Three quantitative metrics, including normalized mean square error (NMSE), the peak signal-to-noise ratio (PSNR), and structural similarity index (SSIM)\cite{SSIM}, were used to measure the reconstruction performance. The smaller NMSE, the larger PSNR and SSIM, indicate a better reconstruction.

\subsection{Experimental Results}
This subsection shows the reconstruction results of the above methods in the In-Distribution and Out-of-Distribution cases. In addition, we show the performance of HFS-SDE with a reduced number of reverse samples iterations. Note that all $x$-fold uniform sampling in this paper refers to full sampling in the $k$-space center while other areas are sampled every $x$ lines. Specifically, a 10-fold uniform undersampling mask implies that one out of every 10 lines is sampled, and 24 lines are fully sampled in the $k$-space center, resulting in an actual acceleration rate of 5.9. Similarly, a 12-fold uniform undersampling indicates sampling one out of every 12 lines, with 22 lines in the $k$-space center fully sampled, achieving an actual acceleration rate of 6.8.

\subsubsection{In-Distribution Experiments}
\begin{table}[!t]
  \caption{\label{tab: In-Distribution expriments}In-Distribution experiments. The average quantitative metrics on the fastMRI knee dataset under 10-fold and 12-fold uniform undersampling.}
  \centering
  \resizebox{\linewidth}{!}{
      \begin{tabular}{c|cccc}
        \hline \hline AF & Method & NMSE (\%) & PSNR (dB) & SSIM (\%) \\
        \hline 
        \multirow{8}{*}{10-fold} & SENSE & 5.04 $\pm$ 4.73 & 25.28 $\pm$ 3.09 & 52.40 $\pm$ 14.00 \\
        & ISTA-Net & {0.66 $\pm$ 0.26} & {33.20 $\pm$ 1.97} & \textbf{87.29 $\pm$ 3.25} \\
        & DeepCascade & 0.75 $\pm$ 0.28 &{32.60 $\pm$ 1.91} & {86.68 $\pm$ 3.18} \\
        & VarNet & 1.16 $\pm$ 0.24 & 31.51 $\pm$ 0.85 & {71.17 $\pm$ 2.74} \\
        & CycleGAN & 0.87 $\pm$ 0.47 & 32.18 $\pm$ 2.03 & 78.89 $\pm$ 7.48 \\
        & VE-SDE & 0.94 $\pm$ 0.49 & 31.76 $\pm$ 1.87 & 78.23 $\pm$ 5.81 \\
        & VP-SDE & 1.14 $\pm$ 1.90 & 32.69 $\pm$ 3.14 & 85.47 $\pm$ 4.20 \\
        & HFS-SDE & \textbf{0.65 $\pm$ 0.26} &  \textbf{33.28 $\pm$ 1.83} & 84.09 $\pm$ 4.11 \\
        \hline 
        \multirow{8}{*}{12-fold} & SENSE & 5.43 $\pm$ 4.59 & 24.81 $\pm$ 2.82 & 51.22 $\pm$ 12.88 \\
        & ISTA-Net & {1.03 $\pm$ 0.63} & {31.47 $\pm$ 1.96} & \textbf{84.57 $\pm$ 3.48} \\
        & DeepCascade & 1.19 $\pm$ 0.64 & 30.80 $\pm$ 1.87 & {83.98 $\pm$ 3.29} \\
        & VarNet & 1.19 $\pm$ 0.21 & 31.36 $\pm$ 1.01 & {70.00 $\pm$ 1.85} \\
        & CycleGAN & 1.07 $\pm$ 0.50 & 31.19 $\pm$ 1.92 & 77.18 $\pm$ 7.36 \\
        & VE-SDE & 1.38 $\pm$ 0.90 & 30.31 $\pm$ 2.06 & 75.19 $\pm$ 5.63 \\
        & VP-SDE & 1.66 $\pm$ 2.44 & 30.92 $\pm$ 3.06 & 82.40 $\pm$ 4.76 \\
        & HFS-SDE & \textbf{0.97 $\pm$ 0.40} & \textbf{31.56 $\pm$ 1.63} & 80.60 $\pm$ 4.32 \\
        \hline \hline
      \end{tabular}
  }
\end{table}

\begin{figure*}
    \centerline{\includegraphics[width=\textwidth]{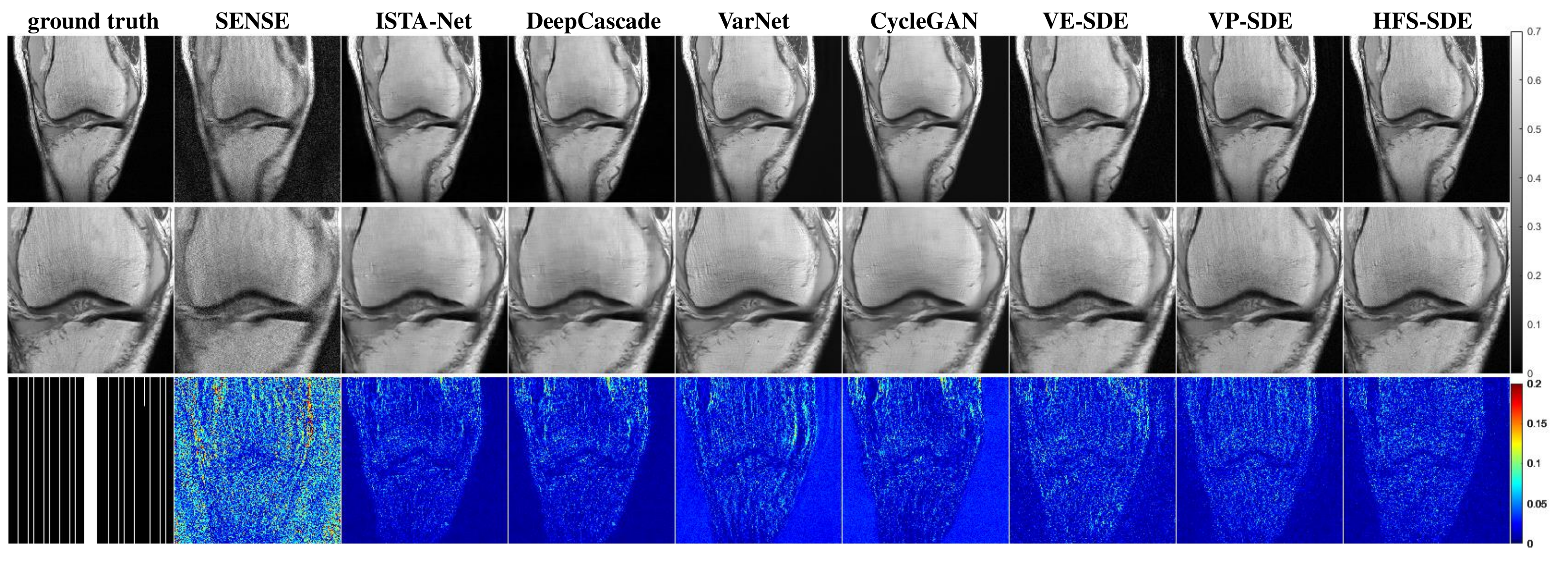}}
    \caption{The reconstruction results of fastMRI multi-coil knee data at uniform undersampling of 10-fold. The first row shows the ground truth and the reconstruction of SENSE, ISTA-Net, DeepCascade, VarNet,cycleGAN, VE-, VP-, and HFS-SDE (ours). The second row shows the enlarged view of the ROI , and the third row shows the error map of the reconstruction. The undersampling mask used for the test is shown in the lower left corner.}
    \label{fig:acc10}
\end{figure*}
In the In-Distribution experiments, the anatomy of the training and testing datasets are the same. Score-based SDEs are unsupervised learning methods in which only the fully sampled images are used for network training without the paired undersampled data. Therefore, no undersampling pattern is needed for network training in VE-, VP-, and HFS-SDEs. For ISTA-Net, the undersampled masks of training and testing are the same.

\begin{figure*}[!t]
    \centering
    \resizebox{\textwidth}{!}{\includegraphics{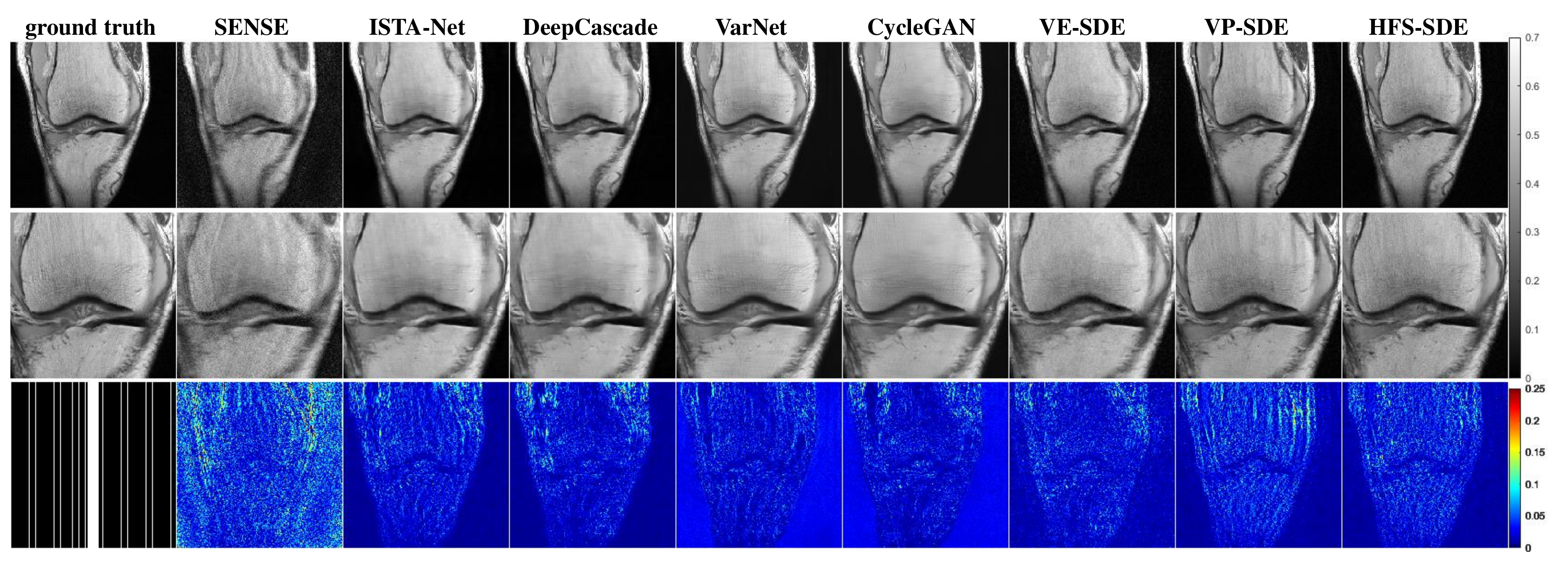}}
    \caption{The reconstruction results of fastMRI multi-coil knee data at uniform undersampling of 12-fold. The first row shows the ground truth and the reconstruction of SENSE, ISTA-Net, DeepCascade, VarNet, CycleGAN, VE-, VP-, and HFS-SDE (ours). The second row shows the enlarged view of the ROI, and the third row shows the error map of the reconstruction. The undersampling mask used for the test is shown in the lower left corner.}
    \label{fig:acc12}
\end{figure*}
The reconstruction results of 10-fold uniform undersampling are shown in \figref{fig:acc10}. The first row shows the reconstruction results of SENSE, ISTA-Net, DeepCascade, VarNet, CycleGAN, VE-SDE, VP-SDE, and HFS-SDE.
The second row displays an enlarged view of the region of interest (ROI). The SENSE reconstruction exhibits noise and aliasing artifacts. Similarly, the reconstructed images by ISTA-Net, DeepCascade, VarNet, and CycleGAN also show aliasing artifacts and blurriness, resulting in the loss of fine structures. Compared to other methods, score-based SDE approaches provide more details. Among these three SDE methods, VE-SDE's reconstruction results lose some details. On the other hand, VP-SDE loses some fine structures in the images. HFS-SDE excels in quantitative metrics, preserving the most realistic high-frequency details and effectively suppressing artifacts.
\figref{fig:acc12} shows the reconstruction results using a 12-fold uniform undersampling mask. The reconstruction images produced by SENSE exhibit severe blurring and aliasing artifacts. ISTA-Net, DeepCascade, VarNet, and CycleGAN suffer a significant loss of details. VE-SDE loses some fine details, and VP-SDE introduces artifacts. In contrast, HFS-SDE achieves the best visual results by preserving the most authentic details. 

The average quantitative metrics for the knee dataset are presented in Table \ref{tab: In-Distribution expriments}. Among all the compared methods, HFS-SDE excels with the best NMSE and PSNR.

\begin{figure*}[!t]
    \centering
    \includegraphics[width=\textwidth]{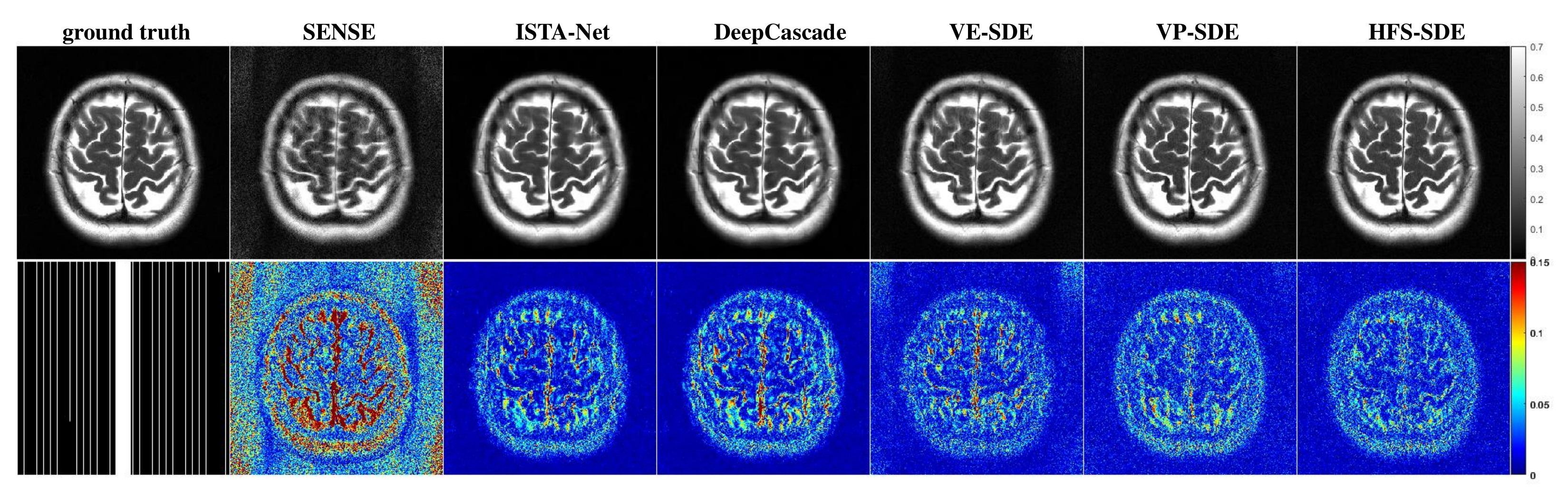}
    \caption{Out-of-Distribution results. The reconstruction results of T2-weighted brain data at uniform undersampling of 12-fold. The first row shows the reconstruction results of SENSE, ISTA-Net, DeepCascade, VE-, VP-, and HFS-SDE. The second row shows the error map of the ROI. The undersampling mask is shown in the lower left corner.}
    \label{fig:data-shift}
\end{figure*}

\subsubsection{Out-of-Distribution Experiments}
We conducted experiments on T2-weighted brain data using models trained on T1-weighted knee data. The models included SENSE, ISTA-Net, DeepCascade, VE-, VP-, and HFS-SDE. In this scenario, the distribution of the testing data differs from that of the training data. The reconstruction results with a 10-fold uniform undersampling are presented in Fig. \ref{fig:data-shift}. ISTA-Net and DeepCascade exhibit a decrease in reconstruction quality, while SDE-based methods maintain high-quality reconstructions. HFS-SDE continues to achieve the best reconstruction results with minimal noise and artifacts. Table .\ref{tab: Out-of-Distribution} displays the quantitative metrics for the test brain dataset.

Additionally, it is essential to emphasize that due to the disparity in the number of parallel channels between the fastMRI knee dataset and the brain dataset, we refrained from evaluating the generalization of the CycleGAN model when handling data with specific channel counts.

\begin{table}[!t]
  \caption{\label{tab: Out-of-Distribution}Out-of-Distribution experiments. The average quantitative metrics on the fastMRI brain dataset under 10-fold uniform undersampling.}
  \centering
  \resizebox{\linewidth}{!}{
      \begin{tabular}{c|ccccc}
        \hline \hline  AF & Methods & NMSE(\%) & PSNR (dB) & SSIM(\%) \\
        \hline
         \multirow{6}{*}{10-fold} & SENSE & 14.31 $\pm$ 13.36 & 22.13 $\pm$ 2.64 & 39.73 $\pm$ 14.78 \\
        & ISTA-Net & 1.65 $\pm$ 1.51 & 31.02 $\pm$ 1.21 & \textbf{85.69 $\pm$ 3.59} \\
        & DeepCascade & 2.03 $\pm$ 1.75 & 30.03 $\pm$ 1.30 & 85.35 $\pm$ 3.15 \\
        & VE-SDE & 2.93 $\pm$ 2.69 & 28.75 $\pm$ 1.77 & 71.35 $\pm$ 9.96 \\
        & VP-SDE & 1.77 $\pm$ 1.01 & 30.16 $\pm$ 1.37 & 72.59 $\pm$ 8.85 \\
        & HFS-SDE & \textbf{1.43 $\pm$ 0.82} & \textbf{31.06 $\pm$ 0.87} & 79.97 $\pm$ 5.14 \\
        \hline \hline
    \end{tabular}
  }
\end{table}

\begin{figure}[!t]
    \centering
    \resizebox{0.5\textwidth}{!}{\includegraphics{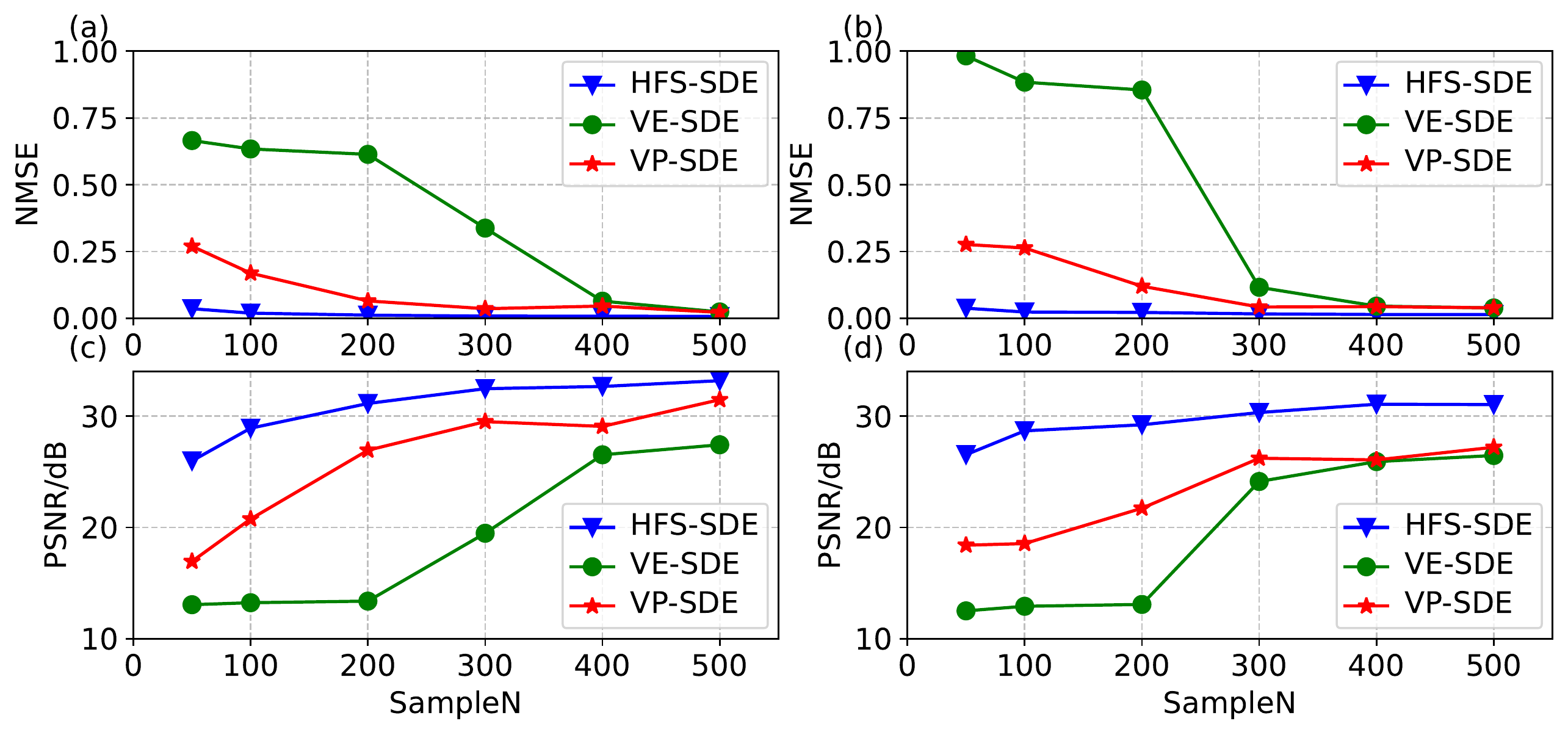}}
    \caption{The quantitative performance-sampling-step diagrams under uniform 10-fold undersampling with sampling steps $N=50$, $100$, $200$, $300$, $400$, and $500$ are presented. Figures (a) and (c) show the NMSE and PSNR metrics for the reconstruction results of fastMRI multi-coil knee data (In-Distribution), while Figures (b) and (d) display the NMSE and PSNR metrics for the reconstruction results of brain data (Out-of-Distribution).}
    \label{fig:sampling_step}
\end{figure}

\subsubsection{Accelerated Sampling}
In the above experiments, the noise scale $N$ of VE-, VP-, and HFS-SDEs was set to $1000$. This implies that $1000$ steps were required in the reverse-time process for sample generation, taking an average of 2.5 minutes to reconstruct a 320 $\times$ 320 MR image on an NVIDIA RTX A100 GPU.

\begin{figure*}[!t]
    \centering
    \includegraphics[width=\textwidth]{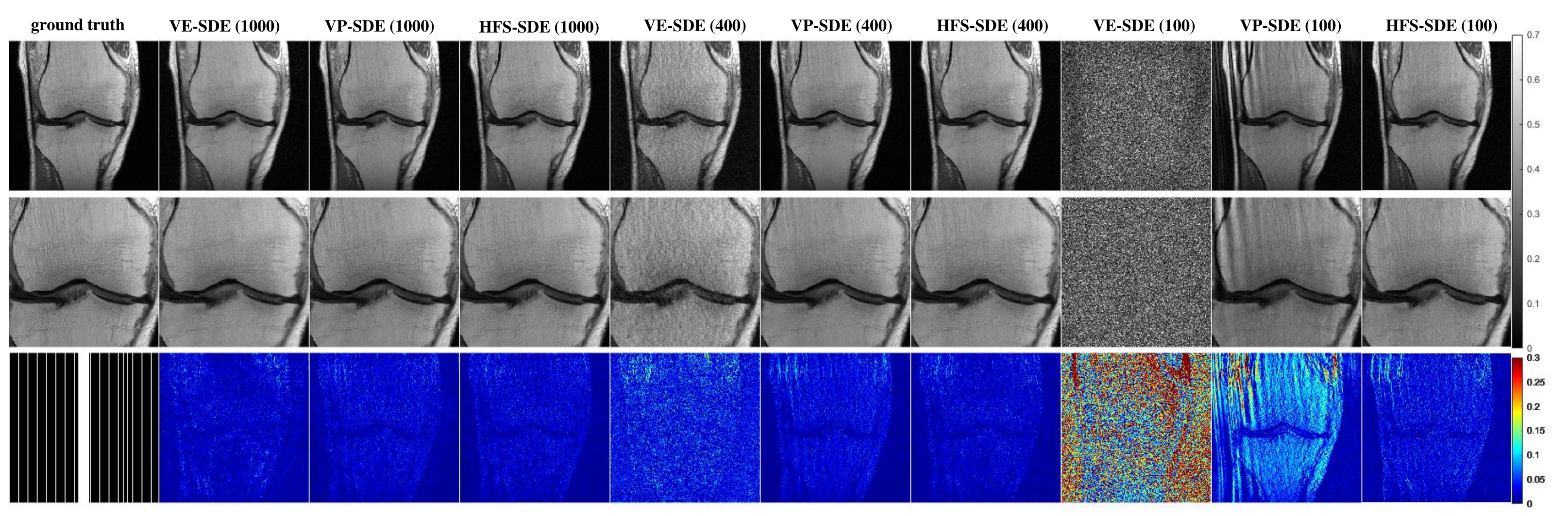}
    \caption{Accelerated Sampling. Reconstruction was performed on the fastMRI multi-coil knee dataset with a 10-fold uniform undersampling, using 100, 400, and 1000 iterations, respectively. The first row presents the ground truth and the reconstructions of VE-, VP-S, and HFS-SDE (our model). The leftmost three columns represent the results of 1000 iterations, the middle three columns represent 400 iterations, and the rightmost three columns represent 100 iterations. The second row displays an enlarged view of the region of interest, indicated by the yellow box in the first row. The third row exhibits the error map of the reconstructions. The undersampling mask used for the test is shown in the lower left corner.}
    \label{fig:Accelerated Sampling}
\end{figure*}

To gain further insights into the acceleration performance of HFS-SDE, we conducted quantitative experiments under varying sampling steps. The model was trained on knee data, and the test data included both knee and brain data with 10-fold uniform undersampling. The sampling steps were set to 50, 100, 200, 300, 400, and 500. Fig. \ref{fig:sampling_step} shows the performance sampling-step diagrams using different SDEs. HFS-SDE converges much faster than VP- and VE-SDE, especially when sampling steps are small. It's worth noting that HFS-SDE achieves relatively low NMSE when the sampling step exceeds 100 on both knee and brain data. Fig. \ref{fig:Accelerated Sampling} showcases the reconstruction results for sampling step sizes of 1000, 400, and 100, respectively. When the number of reverse sampling iterations was reduced to 100, VE-SDE suffered a loss of structural information, and VP-SDE's reconstructed images exhibited severe aliasing artifacts. However, the results of HFS-SDE with $N=100$ were visually comparable to those with $N=1000$, even though the quantitative metrics were slightly lower than those with $N=1000$. When the number of reverse sampling steps was reduced to 400, VE-SDE exhibited a degree of noise, while VP-SDE showed minor aliasing artifacts along the edges. HFS-SDE's performance remained on par with the results at $N=1000$. Additionally, we conducted a time analysis of the reconstruction process for various methods on a single slice, with testing conducted on an NVIDIA RTX A100 GPU. The results are presented in Table \ref{tab: runtime_table}.

For scenarios requiring low noise in reconstruction results, we think $N=400$ is an appropriate choice, as the performance of HFS-SDE is comparable to $N=1000$, effectively suppressing artifacts and noise. For scenarios where reconstruction time is a priority and minor noise is tolerable, $N=100$ is a suitable choice. Although HFS-SDE exhibits slight noise with 100 steps, it still effectively suppresses artifacts and successfully reconstructs structures.

\begin{table*}[h]
  \caption{\label{tab: runtime_table} The time required to reconstruct a slice for each method. For SDE methods, $N=1000$.}
  \centering
  \begin{tabular}{c|cccccccccc}
    \hline\hline
    Method & SENSE & ISTA-Net & DeepCascade & VarNet & CycleGAN & VE-SDE & VP-SDE & HFS-SDE 
    & HFS-SDE(400) & HFS-SDE(100) \\
    \hline
    Time (s) & 1.04 & 0.40 & 0.16 & 0.23 & 0.08 & 104.08 & 122.09 & 98.59 &39.09 & 9.86 \\
    \hline\hline
  \end{tabular}
\end{table*}

\begin{table}[!t]
  \caption{\label{tab:statistic}The results of a statistical hypothesis test on the fastMRI knee dataset. The $t/z$ represents the $t-$ and $z-$variable in the paired samples $T$-test and Wilcoxon Signed Ranks test, respectively, with the corresponding ‘value’ in the value column. * and *** represent a statistical significance level of 0.05 and 0.001, respectively. }
  \centering
  \begin{tabular}{c|ccccc}
    \hline \hline
    \multirow{2}{*}{paired} & \multirow{2}{*}{Metrics} & \multirow{2}{*}{$t/z$} & \multirow{2}{*}{value} & \multirow{2}{*}{p} \\
    & & & & \\
    \hline
    \multirow{3}{*}{HFS-VE} & NMSE & $z$ & 7.86 & 0.000*** \\
           & PSNR & $t$ & 14.56 & 0.000*** \\
           & SSIM & $t$ & 21.38 & 0.000*** \\
    \hline
    \multirow{3}{*}{HFS-VP} & NMSE & $z$ & 0.04 & 0.967\hspace{1.5em} \\
           & PSNR & $t$ & 2.22 & 0.028*\hspace{1em}  \\
           & SSIM & $t$ & -5.66 & 0.000*** \\
    \hline \hline
  \end{tabular}
\end{table}

\subsubsection{Statistical Hypothesis}
We conducted a statistical hypothesis test on the fastMRI knee dataset to demonstrate the significance of our experimental results. Among the three metrics, NMSE did not follow a normal distribution, and a paired Wilcoxon Signed Ranks test (Wilcoxon Signed Ranks test) was employed. PSNR and SSIM followed normal distributions and were analyzed using paired samples $T$-test. The results (shown in Table ~\ref{tab:statistic}) demonstrate that HFS-SDE exhibits statistical significance in terms of PSNR and SSIM compared to VP-SDE ($p-value < 0.05$) and in all three metrics compared to VP-SDE ($p-value < 0.001$).

\section{Discussion}\label{discussion}
In this study, the HFS-SDE was proposed for solving the inverse problem in accelerated MRI based on the diffusion model. We have demonstrated that diffusion in the high-frequency space effectively improves the stability of VP-SDE. Meanwhile, HSF-SDE reduces the time required for reverse sampling and maintains the reconstruction quality by utilizing the low-frequency information of MR images.

\subsection{Extensions of HSF-SDE}
The idea of diffusion in high-frequency space can be extended to other SDEs, such as VE- and sub-VP-SDEs. The corresponding forward SDEs are: 
\begin{equation}
    \left\{\begin{array}{l}
    \mathrm{d} \mathbf{x}=\sqrt{\frac{\mathrm{d}\left[\sigma^2(t)\right]}{\mathrm{d} t}}\boldsymbol{\mathcal{F}}_h \mathrm{d} \mathbf{w}\\
    \mathrm{d} \mathbf{x}=-\frac{1}{2} \beta(t) \boldsymbol{\mathcal{F}}_h(\mathbf{x}) \mathrm{d} t+\sqrt{\beta(t)\left(1-e^{-2 \int_0^t \beta(s) \mathrm{d} s}\right)} \boldsymbol{\mathcal{F}}_h\mathrm{d} \mathbf{w}
    \end{array}\right..
\end{equation}
Then MR reconstructions using these two SDEs will start from the low-resolution image generated from the $k$-space center instead of pure Gaussian noise, hence reducing sampling iterations in the reverse process. One major drawback of fast MRI is the loss of details.  Diffusion in HFS may help recover high-frequency details on MR images as well.  

\subsection{Ablation Studies}
We conducted an ablative study on the selection of the center size $n_l$   in the mask $M_l$. Specifically, $n_l$ = 2, 8, 16, 24, and 32 were chosen to train the model of HFS-SDE. The results of the 10- and 12-fold uniform undersampling knee data are presented in Table ~\ref{tab: ablation: nl}. HFS-SDE achieves similar performance when $n_l$ is set within the range of 8 to 24, and the optimal performance of HFS-SDE was achieved when $n_l = 16$. A center size that is too large or too small leads to a degradation of HFS-SDE’s performance. Additionally, a center area that is too small may also affect the performance of accelerated sampling. Therefore, we used $n_l = 16$ in this study.

\begin{figure*}[!t]
    \centering
    \resizebox{\textwidth}{!}{\includegraphics{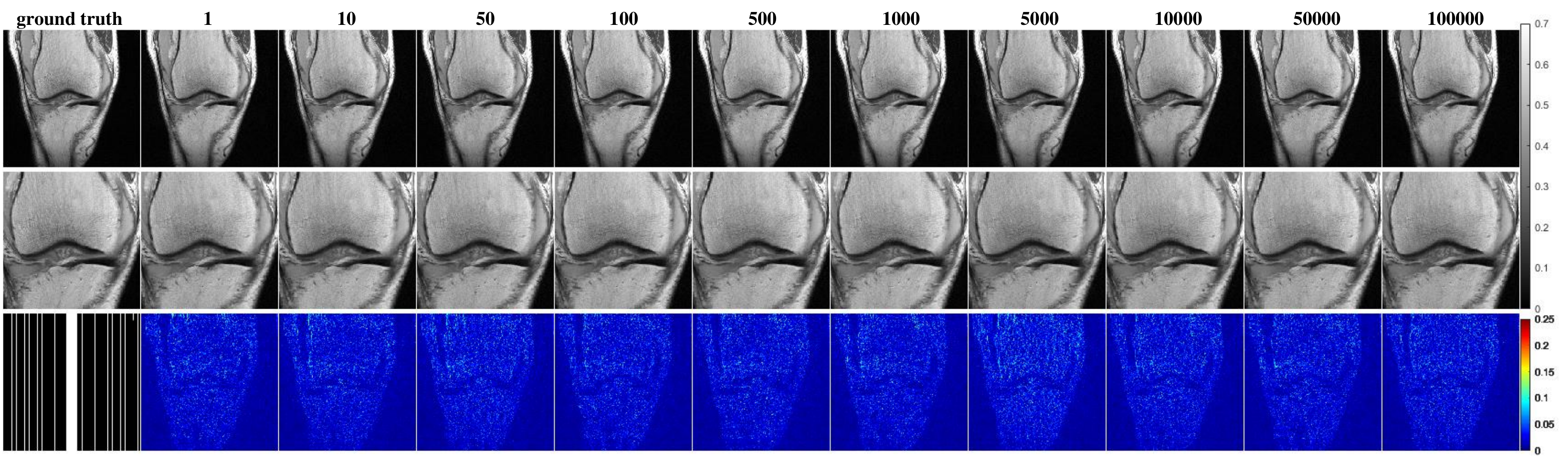}}
    \caption{Random seed experiments in HFS-SDE. The reconstruction results of fastMRI multi-coil knee data at uniform undersampling of 10-fold. The first row shows the ground truth and the reconstruction of different seeds. The second row displays the error map of reconstructions. The undersampling mask used for the test is shown in the lower left corner.}
    \label{fig:knee_seed}
\end{figure*}

We also conducted experiments with various random noise initializing seeds (seeds: 1, 10, 50, 100, 500, 1000, 5000, 10000, 50000, and 100000) during the reconstruction process to understand the robustness and reliability of our HFS-SDE. 
The reconstruction results of fastMRI multi-coil knee data with different random seeds under 10-fold uniform undersampling are presented in Fig. \ref{fig:knee_seed}. It can be observed that there are no significant differences in the reconstruction results across various random seeds, demonstrating the robustness and reliability of HFS-SDE. And the average quantitative metric on the fastMRI knee dataset under 10-fold uniform undersampling is presented in Table~\ref{tab: Seed Experiments Combined}.

\begin{table}[!t]
  \caption{\label{tab: Seed Experiments Combined}Ablation studies on random seed. The average quantitative metrics on the fastMRI knee dataset under 10-fold uniform undersampling with different random seeds.}
  \centering
  \resizebox{\linewidth}{!}{
      \begin{tabular}{c|cccc}
        \hline \hline Seed & NMSE (\%) & PSNR (dB) & SSIM (\%) \\
        \hline 
        1 & 0.66 $\pm$ 0.29 & 33.28 $\pm$ 1.82 & \textbf{84.12 $\pm$ 4.31} \\
        10 & 0.67 $\pm$ 0.30 & 33.22 $\pm$ 1.85 & 84.10 $\pm$ 4.37 \\
        50 & 0.65 $\pm$ 0.28 & 33.30 $\pm$ 1.83 & 84.10 $\pm$ 4.35 \\
        100 & 0.66 $\pm$ 0.27 & 33.24 $\pm$ 1.75 & 84.08 $\pm$ 4.33 \\
        500 & 0.67 $\pm$ 0.29 & 33.17 $\pm$ 1.83 & 84.06 $\pm$ 4.39 \\
        1000 & 0.67 $\pm$ 0.30 & 33.21 $\pm$ 1.88 & 84.02 $\pm$ 4.40 \\
        5000 & \textbf{0.65 $\pm$ 0.27} & \textbf{33.33 $\pm$ 1.79} & 84.11 $\pm$ 4.38 \\
        10000 & 0.66 $\pm$ 0.28 & 33.23 $\pm$ 1.82 & 84.07 $\pm$ 4.33 \\
        50000 & 0.66 $\pm$ 0.30 & 33.27 $\pm$ 1.86 & 84.08 $\pm$ 4.44 \\
        100000 & 0.67 $\pm$ 0.29 & 33.23 $\pm$ 1.84 & 84.05 $\pm$ 4.44 \\
        \hline \hline
      \end{tabular}
  }
\end{table}

\subsection{Experiment with Gaussian undersampling Mask}
\begin{figure*}[!t]
    \centering
    \includegraphics[width=\textwidth]{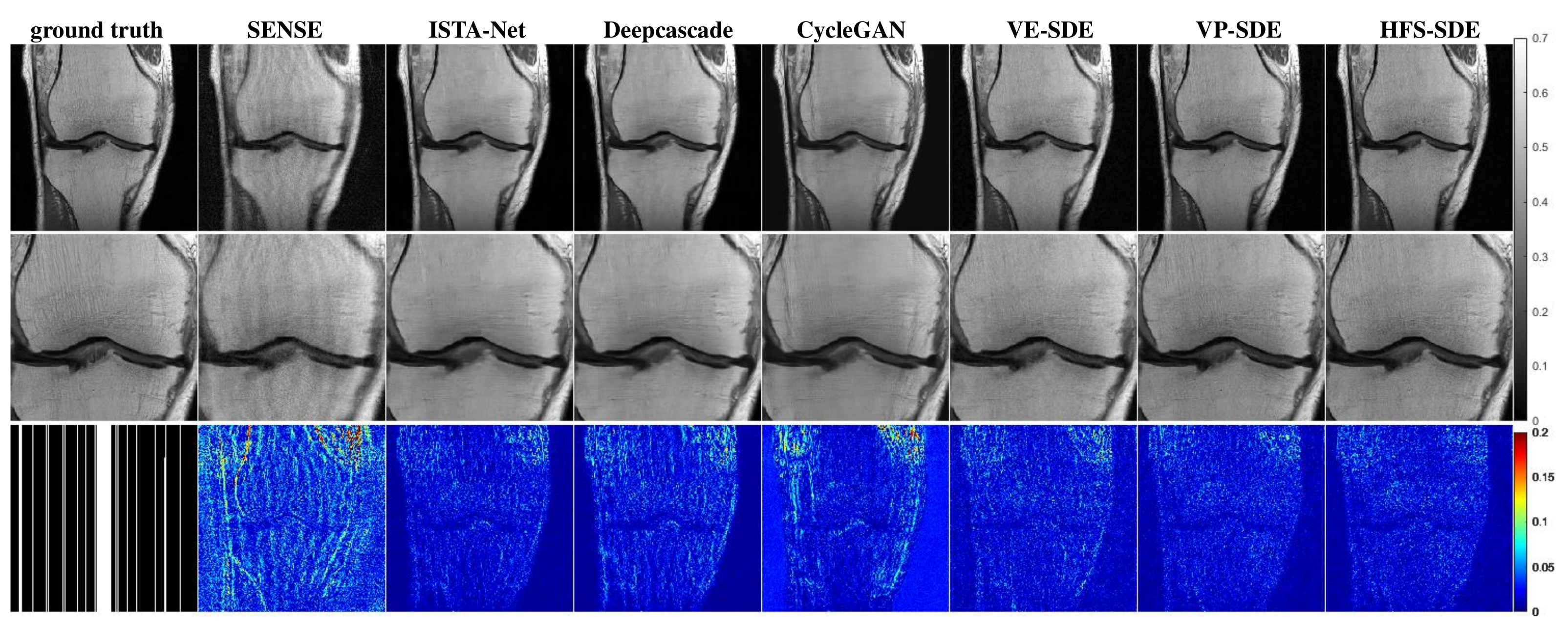}
    \caption{Gaussian mask experiment. The reconstruction results of fastMRI multi-coil knee data at random undersampling of 6-fold. The first row shows the ground truth and the reconstruction of SENSE, Deepcascade,VE-, VP-, and HFS-SDE (ours). The second row shows the enlarged view of the ROI, and the third row shows the error map of the reconstruction. The undersampling mask used for the test is shown in the lower left corner.}
    \label{fig:random_uniform_knee}
\end{figure*}

Since the mode of undersampling mask may affect the reconstruction performance, we performed the experiments using a Gaussian undersampling mask with an actual acceleration rate = 5.9 on the knee dataset. The reconstruction images are shown in Fig. \ref{fig:random_uniform_knee} with quantitative metrics presented in Table .\ref{tab: random mask}. Similar conclusions can be drawn from the results using Gaussian undersampling mask, as from Fig. \ref{fig:acc10}, that HFS-SDE still outperforms other methods. The performance of most DL-based methods slightly degrades when using a Gaussian mask relative to a uniform mask, which is consistent with the previous studies\cite{levine2017fly, athalye2015parallel, haldar2016p}.  

\subsection{Harsh Condition Experiment}
\begin{figure*}[h]
    \centering
    \resizebox{\textwidth}{!}{\includegraphics{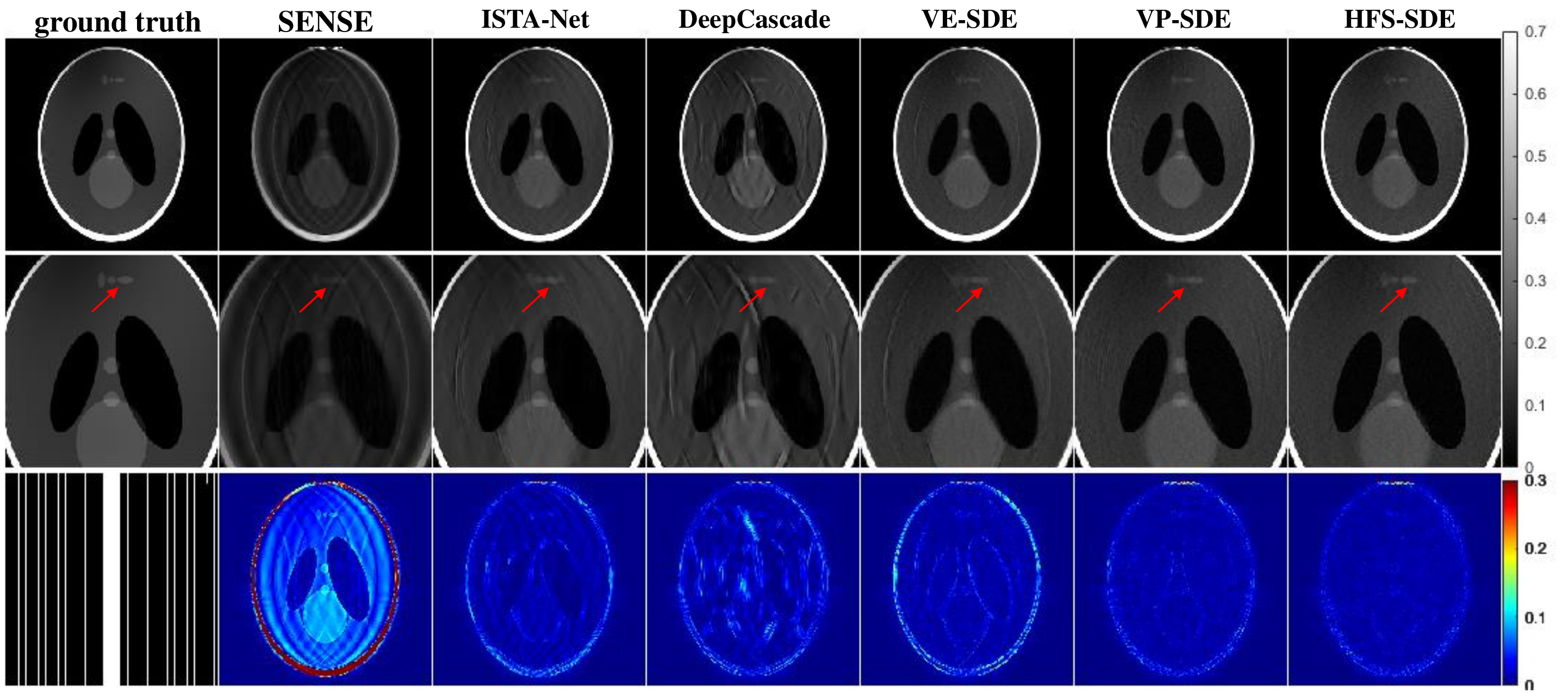}}
    \caption{Phantom simulation. The reconstruction results of the phantom at uniform undersampling of 10-fold. The first row shows the ground truth and the reconstruction of SENSE, ISTA-Net, Deepcascade, VE-, VP-, and HFS-SDE (ours). The second row shows the enlarged view of the ROI (The red arrows point to the details), and the third row shows the error map of the reconstruction. The undersampling mask used for the test is shown in the lower left corner.}
    \label{fig:photom_uniform10}
\end{figure*}

We conducted a harsh condition experiment using a simulated phantom with large black and grey areas. The models of SENSE, ISTA-Net, DeepCascade, VE-, VP-, and HFS-SDE trained on the knee data were employed. The distribution of the training data was different from that of the simulated phantom. Fig. \ref{fig:photom_uniform10} illustrates the reconstruction results under 10-fold uniform undersampling. Under such a challenging condition, SENSE, ISTA-Net, and DeepCascade exhibit severe aliasing artifacts. In contrast, the SDE-based models have no obvious artifacts. Among all compared methods, HFS-SDE achieves the most outstanding results, preserving the most details (indicated by the red arrows).

$\lambda_2$ is the coefficient in the Corrector algorithm used to control data consistency, which has an impact on the quality of reconstruction.
If $\lambda_2$ is not within the appropriate range, the SDE-based models might yield noticeable errors in phantom reconstructions. However, SDE-based models can still effectively reduce artifacts in the region of interest, even with suboptimal parameters. On the other hand, while methods such as ISTA-Net and DeepCascade do not exhibit noticeable errors, their performance in suppressing artifacts is subpar. For specific hyperparameter selection, please refer to our code.

\begin{table}[!t]
  \caption{\label{tab: ablation: nl}Ablation studies on the selection of $\mathbf{M}_l$. The average quantitative metrics on the fastMRI knee dataset under 10-fold and 12-fold uniform undersampling.}
  \centering
  \resizebox{\linewidth}{!}{
      \begin{tabular}{c|cccc}
        \hline \hline AF & $n_l$ & NMSE(\%) & PSNR (dB) & SSIM(\%) \\
        \hline 
        \multirow{5}{*}{10-fold}
        & 2 & 1.21 $\pm$ 1.92 & 32.04 $\pm$ 2.91 & 80.99 $\pm$ 6.26 \\
        & 8 & \textbf{0.64 $\pm$ 0.27} & \textbf{33.35 $\pm$ 1.83} & \textbf{84.73 $\pm$ 4.11} \\
        & 16 & 0.65 $\pm$ 0.26 & 33.28 $\pm$ 1.83 & 84.09 $\pm$ 4.11 \\
        & 24 & 0.72 $\pm$ 0.49 & 33.06 $\pm$ 1.97 & 82.99 $\pm$ 5.74 \\
        & 32 & 1.10 $\pm$ 0.59 & 31.15 $\pm$ 1.83 & 76.06 $\pm$ 7.36 \\
        \hline
        \multirow{5}{*}{12-fold}& 2 & 1.34 $\pm$ 1.23 & 30.74 $\pm$ 2.11 & 78.47 $\pm$ 5.46 \\
        & 8 & 1.18 $\pm$ 0.68 & 30.90 $\pm$ 1.79 & 77.99 $\pm$ 5.71 \\
        & 16 & \textbf{0.97 $\pm$ 0.40} & \textbf{31.56 $\pm$ 1.63} & \textbf{80.60 $\pm$ 4.32} \\
        & 24 & 1.11 $\pm$ 0.73 & 31.21 $\pm$ 1.98 & 79.01 $\pm$ 6.50 \\
        & 32 & 1.93 $\pm$ 1.11 & 28.81 $\pm$ 1.70 & 70.62 $\pm$ 8.13 \\
        \hline \hline
    \end{tabular}
  }
\end{table}

\subsection{HFS with Multiple Diffusion Processes}
Based on the frequency operators, the image is decomposed into high- and low-frequency parts, where different diffusion processes can be applied. VP-SDE can be treated as applying Gaussian noise in both low- and high-frequency. Additionally, we can perform the diffusion process with Rician noise for the low-frequency part and Gaussian noise for the high-frequency part of the image (refer to HFS-Rician). However, HFS-Rician has a degraded performance with NMSE = 2.25±1.56 \%, PSNR = 28.49±2.38 dB, and SSIM = 76.11±6.08 \%, compared with HFS-SDE. The main reason is adding noise to the low-frequency part results in a compromised consistency between the generated samples and the acquired $k$-space data. This finding also confirms the motivation of this study.

\subsection{Data Harmonization}
Data harmonization should be considered when applying the proposed method for the usage of multicenter data deployment. It means data from various sources\cite{nan2022data} can be analyzed and compared within a unified framework, thereby alleviating disparities and discrepancies from different data sources\cite{nan2022data}. For MR images from multicenter, data differs on noise level, image size, grey intensity, etc, due to the changes in imaging parameters and vendors.   In our study, we adopted an image preprocessing procedure to enhance data harmonization. Firstly, we performed image resampling to obtain a consistent image size, facilitating comparisons and analysis in subsequent steps. Subsequently, we normalized the image intensity using the standard deviation normalization method. This standardization process ensured consistent scales and ranges for the input data during model training and evaluation. Although the fast-MRI dataset employed in this study is a single-center dataset, we believe this procedure also works for multi-center data to improve the reproducibility and generalization of reconstruction methods. 

\begin{table}[!t]
  \caption{\label{tab: random mask} Gaussian mask experiments. The average quantitative metrics on the fastMRI knee dataset under 6-fold random undersampling}
  \centering
  \resizebox{\linewidth}{!}{
      \begin{tabular}{c|cccc}
        \hline \hline Method & NMSE (\%) & PSNR (dB) & SSIM (\%) \\
        \hline 
         SENSE & 4.64 $\pm$ 3.94 & 25.51 $\pm$ 2.68 & 56.59 $\pm$ 12.86 \\
        ISTA-Net & 0.76 $\pm$ 0.37 & 32.69 $\pm$ 2.00 & \textbf{86.89 $\pm$ 3.12} \\
        DeepCascade & 0.91 $\pm$ 0.35 & 31.82 $\pm$ 1.85 & 86.10 $\pm$ 3.02 \\
        CycleGAN & 1.42 $\pm$ 0.52 & 29.87 $\pm$ 1.68 & 75.34 $\pm$ 7.01 \\
        VE-SDE & 0.94 $\pm$ 0.49 & 31.82 $\pm$ 1.87 & 78.64 $\pm$ 5.53 \\
        VP-SDE & 2.07 $\pm$ 4.64 & 31.36 $\pm$ 3.69 & 81.31 $\pm$ 6.90 \\
        HFS-SDE & \textbf{0.70 $\pm$ 0.28} & \textbf{32.98 $\pm$ 1.86} & 83.80 $\pm$ 4.01 \\
        \hline \hline
      \end{tabular}
  }
\end{table}

\subsection{Limitations}
There exist several U-Net architectures to estimate the score function in the diffusion model, with different skip connections, residual blocks, etc. The choice of the U-Net architecture affects the performance of the diffusion model. In this study, we just used the \texttt{ddpm} provided by Song et al. \cite{score-based-SDE} and did not test other network architectures since our study aims to demonstrate the advantage of diffusion in high-frequency space. For a fair comparison, all three SDE-based reconstruction methods used the same network architecture in the experiments. A more complicated network may further improve the performance of the proposed method. The test dataset is small in this study. Although we have shown the advantages of HSF-SDE on this test dataset, testing the performance of HSF-SDE on a larger dataset is still needed and will be investigated in our future work.  

\section{Conclusions}\label{conclusions}
A new SDE was proposed for fast MR reconstruction that restricts the diffusion process in high-frequency space. It ensures determinism in the fully sampled low-frequency regions and accelerates the sampling procedure of reverse diffusion. Our experiments have demonstrated that HFS-SDE reconstruction method outperforms the parallel imaging, supervised deep learning, and existing VE- and VP-SDEs-based methods in terms of reconstruction accuracy.

\section*{Acknowledgement}
This study was supported in part by the National Natural Science Foundation of China under grant nos. 62322119, 12226008, 62125111, U1805261, 62106252, 62206273, 62201561, National Key R\&D Program of China nos. 2020YFA0712200, 2021YFF0501402, the Guangdong Basic and Applied Basic Research Foundation no. 2023B1212060052, Shenzhen Science and Technology Program under grant no. RCYX20210609104444089, JCYJ20220818101205012.

\appendix
\subsection{The Forward Diffusion Process of HFS-SDE}\label{appendix:A}
This section presents the detailed derivation of the forward diffusion process for HFS-SDE. Assuming that there are $N$ noise scales in the forward diffusion process, the following discrete Markov chain constructs the forward diffusion process,
\begin{multline}
    \mathbf{x}_{i}=\boldsymbol{\mathcal{F}}_l(\mathbf{x}_{i-1}) + \sqrt{1-\beta_{i}} \boldsymbol{\mathcal{F}}_h(\mathbf{x}_{i-1})+\sqrt{\beta_{i}} \boldsymbol{\mathcal{F}}_h(\mathbf{z}_{i-1}), \\ \quad i=1, \cdots, N,
    \label{Appendix HFS discrete Markov}
\end{multline}
where $\{\beta_{1}, \beta_{2}, \cdots, \beta_{N}\}$ are given coefficients to control the noise level and $\mathbf{z}_{i-1} \sim \mathcal{N}(\mathbf{0}, \mathbf{I})$. By introducing auxiliary variables $\bar{\beta}_{i}=N \beta_{i}$, \eqnref{Appendix HFS discrete Markov} can be rewritten as
\begin{multline}
    \mathbf{x}_{i}=\boldsymbol{\mathcal{F}}_l(\mathbf{x}_{i-1}) + \sqrt{1-\frac{\bar{\beta}_{i}}{N}} \boldsymbol{\mathcal{F}}_h(\mathbf{x}_{i-1})+\sqrt{\frac{\bar{\beta}_{i}}{N}} \boldsymbol{\mathcal{F}}_h(\mathbf{z}_{i-1}), \\ \quad i=1, \cdots, N.
\end{multline}
When $N \rightarrow \infty$ and $t \in [0, 1]$, $\{\bar{\beta}_{i}\}_{i=1}^N$ can be written as $\{\beta(t)\}_{t=0}^1$, $\{\mathbf{z_i}_{i=1}^N\}$ become $\{\mathbf{z}(t)\}_{t=0}^1$ and $\{\mathbf{x_i}_{i=1}^N\}$ become a continuous process $\{\mathbf{x}(t)\}_{t=0}^1$. Let $\Delta t=\frac{1}{N} \rightarrow 0$,
\begin{multline}
\begin{aligned}
    &~~~\mathbf{x}(t+\Delta t)\\ &=\boldsymbol{\mathcal{F}}_l(\mathbf{x}(t)) + \sqrt{1-\beta(t+\Delta t) \Delta t} \boldsymbol{\mathcal{F}}_h(\mathbf{x}(t))\\&~~~+\sqrt{\beta(t+\Delta t) \Delta t} \boldsymbol{\mathcal{F}}_h(\mathbf{z}(t)) \\
    &\approx \boldsymbol{\mathcal{F}}_l(\mathbf{x}(t)) + \boldsymbol{\mathcal{F}}_h(\mathbf{x}(t)) -\frac{1}{2} \beta(t+\Delta t) \Delta t \boldsymbol{\mathcal{F}}_h(\mathbf{x}(t))\\&~~~+\sqrt{\beta(t+\Delta t) \Delta t} \boldsymbol{\mathcal{F}}_h(\mathbf{z}(t)) \\
    &\approx \mathbf{x}(t)-\frac{1}{2} \beta(t) \Delta t \boldsymbol{\mathcal{F}}_h(\mathbf{x}(t))+\sqrt{\beta(t) \Delta t} \boldsymbol{\mathcal{F}}_h(\mathbf{z}(t)).
\end{aligned}
\label{HFS-SDE converge}
\end{multline}
\eqnref{HFS-SDE converge} converges to
\begin{equation}
    \mathrm{d} \mathbf{x}=-\frac{1}{2} \beta(t) \boldsymbol{\mathcal{F}}_h(\mathbf{x}) \mathrm{d} t+\sqrt{\beta(t)} \boldsymbol{\mathcal{F}}_h\mathrm{d}\mathbf{w}.
\end{equation}

\subsection{Estimate Score for HFS-SDE} \label{appendix:B}
The score model $\mathbf{s}_{\boldsymbol{\theta}}(\mathbf{x}(t), t)$ can be trained via score matching to estimate score function $\log p_{0 t}(\mathbf{x}(t) \mid \mathbf{x}(0))$:
\begin{multline}
        \boldsymbol{\theta}^{*}=\underset{\boldsymbol{\theta}}{\arg \min } \mathbb{E}_{t}\big\{\lambda(t) \mathbb{E}_{\mathbf{x}(0)} \mathbb{E}_{\mathbf{x}(t) \mid \mathbf{x}(0)}\big[\big\|\mathbf{s}_{\boldsymbol{\theta}}(\mathbf{x}(t), t)\\-\nabla_{\mathbf{x}(t)} \log p_{0 t}(\mathbf{x}(t) \mid \mathbf{x}(0))\big\|_{2}^{2}\big]\big\},
    \label{estimate score}
\end{multline}
where $\lambda(t)$ is the positive weighting function. By \eqnref{appendix mean and covariance}, \eqnref{estimate score} can be rewritten as:
\begin{multline}
    \boldsymbol{\theta}^{*}=\underset{\boldsymbol{\theta}}{\arg \min } \mathbb{E}_{t}\big\{\lambda(t) \mathbb{E}_{\mathbf{x}(0)} \mathbb{E}_{\mathbf{x}(t) \mid \mathbf{x}(0)}\big[\big\|\mathbf{z}+\\\sqrt{\big(1-e^{-\frac{1}{2} t^{2}\left(\bar{\beta}_{\max }-\bar{\beta}_{\min }\right)-t \bar{\beta}_{\min }}\big)\boldsymbol{\mathcal{F}}_h}(\mathbf{s}_{\boldsymbol{\theta}}(\mathbf{x}(t), t))\big\|_{2}^{2}\big]\big\},
\end{multline}
where $\mathbf{z} \sim \mathcal{N}(\mathbf{0}, \mathbf{I})$. Since $\boldsymbol{\mathcal{F}}_h(\cdot)^{\frac{1}{2}}=\boldsymbol{\mathcal{F}}_h(\cdot)$, we finally trained the network with following loss:
\begin{multline}
    \boldsymbol{\theta}^{*}=\underset{\boldsymbol{\theta}}{\arg \min } \mathbb{E}_{t}\big\{\lambda(t) \mathbb{E}_{\mathbf{x}(0)} \mathbb{E}_{\mathbf{x}(t) \mid \mathbf{x}(0)}\big[\big\|\boldsymbol{\mathcal{F}}_h(\mathbf{z})+\\\sqrt{\big(1-e^{-\frac{1}{2} t^{2}\left(\bar{\beta}_{\max }-\bar{\beta}_{\min }\right)-t \bar{\beta}_{\min }}\big)}\boldsymbol{\mathcal{F}}_h(\mathbf{s}_{\boldsymbol{\theta}}(\mathbf{x}(t), t))\big\|_{2}^{2}\big]\big\}.
\end{multline}
\subsection{Proof for Theorem \ref{thm:1}}
\label{app: Proof}
Firstly, the generator for HFS-SDE (\ref{reverse sde}) is expressed as:
\begin{equation}
\begin{aligned}
\mathcal{L}_{\text{high}
}f(\mathbf{x}(t))=-\dfrac{1}{2}(&\beta(t) \boldsymbol{\mathcal{F}}_h(\mathbf{x}(t)+2\nabla_{\mathbf{x}} \log p_{t}(\mathbf{x}(t) \mid \mathbf{y}))\cdot\nabla\\&-\beta(t)\boldsymbol{\mathcal{F}}_h\boldsymbol{\mathcal{F}}_h^*:\nabla\nabla^*)f(\mathbf{x}(t)),
\end{aligned} 
\end{equation}
where $\mathbf{a} \cdot \mathbf{b} := \mathbf{a}^* \mathbf{b}$ and  $\mathbf{A}: \mathbf{B}:=  \operatorname{tr}\left(\mathbf{A}^* \mathbf{B}\right)$. 
The generator for the full space version of (\ref{reverse sde}) takes the form:
\begin{equation}
\begin{aligned}
\mathcal{L}_{\text{full}}f(\mathbf{x}(t))=-\dfrac{1}{2}(&\beta(t) (\mathbf{x}(t)+2\nabla_{\mathbf{x}} \log p_{t}(\mathbf{x}(t) \mid \mathbf{y}))\cdot\nabla\\&-\beta(t)\boldsymbol{\mathcal{I}}:\nabla\nabla^* )f(\mathbf{x}(t)). 
\end{aligned} 
\end{equation}
Following \cite{NIPS2015_af473271}, the weak convergence of discrete HFS-SDE and its full space version is bounded by
\begin{equation}
\begin{aligned}
& \mathbb{E}\left[\frac{1}{N} \sum_i \phi\left(\mathbf{x}_{i}\right)-\bar{\phi}\right] \\
= & \frac{1}{N h}\left(\mathbb{E}\left[\psi\left(\mathbf{x}_{0}\right)\right]-\psi\left(\mathbf{x}_T\right)\right)-\sum_{k=2}^K \frac{h^{k-1}}{k ! N} \sum_{l=1}^N \mathbb{E}\left[\mathcal{L}_{(\cdot)}^k \psi\left(\mathbf{x}_{N-l}\right)\right]\\&+O\left(h^K\right)
\end{aligned} 
\end{equation}
where $\mathcal{L}_{(\cdot)} $ denotes $\mathcal{L}_{\text{high}}$ or $\mathcal{L}_{\text{full}}$.
Due to $\|\mathcal{L}_{\text{high}}\| \leq \|\mathcal{L}_{\text{full}}\|$, the upper bound of the second term on the right-hand side of the discrete HFS-SDE equation is smaller than that of its full-space version. Therefore, HFS-SDE exhibits a smaller weak convergence upper bound than its full-space version.

\bibliographystyle{IEEEtran}
\bibliography{refs}

\begin{thebibliography}{10}
\providecommand{\url}[1]{#1}
\csname url@samestyle\endcsname
\providecommand{\newblock}{\relax}
\providecommand{\bibinfo}[2]{#2}
\providecommand{\BIBentrySTDinterwordspacing}{\spaceskip=0pt\relax}
\providecommand{\BIBentryALTinterwordstretchfactor}{4}
\providecommand{\BIBentryALTinterwordspacing}{\spaceskip=\fontdimen2\font plus
\BIBentryALTinterwordstretchfactor\fontdimen3\font minus
  \fontdimen4\font\relax}
\providecommand{\BIBforeignlanguage}[2]{{%
\expandafter\ifx\csname l@#1\endcsname\relax
\typeout{** WARNING: IEEEtran.bst: No hyphenation pattern has been}%
\typeout{** loaded for the language `#1'. Using the pattern for}%
\typeout{** the default language instead.}%
\else
\language=\csname l@#1\endcsname
\fi
#2}}
\providecommand{\BIBdecl}{\relax}
\BIBdecl

\bibitem{haldar2010compressed}
J.~P. Haldar, D.~Hernando, and Z.-P. Liang, ``Compressed-sensing mri with
  random encoding,'' \emph{IEEE Transactions on Medical Imaging}, vol.~30,
  no.~4, pp. 893--903, 2010.

\bibitem{jin2016general}
K.~H. Jin, D.~Lee, and J.~C. Ye, ``A general framework for compressed sensing
  and parallel mri using annihilating filter based low-rank hankel matrix,''
  \emph{IEEE Transactions on Computational Imaging}, vol.~2, no.~4, pp.
  480--495, 2016.

\bibitem{majumdar2015improving}
A.~Majumdar, ``Improving synthesis and analysis prior blind compressed sensing
  with low-rank constraints for dynamic mri reconstruction,'' \emph{Magnetic
  Resonance Imaging}, vol.~33, no.~1, pp. 174--179, 2015.

\bibitem{pruessmann1999sense}
K.~P. Pruessmann, M.~Weiger, M.~B. Scheidegger, and P.~Boesiger, ``Sense:
  sensitivity encoding for fast mri,'' \emph{Magnetic Resonance in Medicine},
  vol.~42, no.~5, pp. 952--962, 1999.

\bibitem{grappa}
M.~A. Griswold, P.~M. Jakob, R.~M. Heidemann, M.~Nittka, V.~Jellus, J.~Wang,
  B.~Kiefer, and A.~Haase, ``Generalized autocalibrating partially parallel
  acquisitions (grappa),'' \emph{Magnetic Resonance in Medicine}, vol.~47,
  no.~6, pp. 1202--1210, 2002.

\bibitem{lustig2010spirit}
M.~Lustig and J.~M. Pauly, ``Spirit: iterative self-consistent parallel imaging
  reconstruction from arbitrary k-space,'' \emph{Magnetic Resonance in
  Medicine}, vol.~64, no.~2, pp. 457--471, 2010.

\bibitem{kt-SLR}
S.~G. Lingala, Y.~Hu, E.~DiBella, and M.~Jacob, ``Accelerated dynamic mri
  exploiting sparsity and low-rank structure: k-t slr,'' \emph{IEEE
  Transactions on Medical Imaging}, vol.~30, no.~5, pp. 1042--1054, 2011.

\bibitem{Low-Rank}
C.~Y. Lin and J.~A. Fessler, ``Efficient dynamic parallel mri reconstruction
  for the low-rank plus sparse model,'' \emph{IEEE Transactions on
  Computational Imaging}, vol.~5, no.~1, pp. 17--26, 2019.

\bibitem{zhao2012image}
B.~Zhao, J.~P. Haldar, A.~G. Christodoulou, and Z.-P. Liang, ``Image
  reconstruction from highly undersampled (k, t)-space data with joint partial
  separability and sparsity constraints,'' \emph{IEEE Transactions on Medical
  Imaging}, vol.~31, no.~9, pp. 1809--1820, 2012.

\bibitem{huang2023deep}
J.~Huang, P.~F. Ferreira, L.~Wang, Y.~Wu, A.~I. Aviles-Rivero, C.-B. Schonlieb,
  A.~D. Scott, Z.~Khalique, M.~Dwornik, R.~Rajakulasingam \emph{et~al.}, ``Deep
  learning-based diffusion tensor cardiac magnetic resonance reconstruction: A
  comparison studies,'' \emph{arXiv preprint arXiv:2304.00996}, 2023.

\bibitem{huang2023vigu}
J.~Huang, A.~I. Aviles-Rivero, C.-B. Sch{\"o}nlieb, and G.~Yang, ``Vigu: Vision
  gnn u-net for fast mri,'' in \emph{2023 IEEE 20th International Symposium on
  Biomedical Imaging (ISBI)}.\hskip 1em plus 0.5em minus 0.4em\relax IEEE,
  2023, pp. 1--5.

\bibitem{huang2022unsupervised}
P.~Huang, C.~Zhang, X.~Zhang, X.~Li, L.~Dong, and L.~Ying, ``Unsupervised deep
  unrolled reconstruction using regularization by denoising,'' \emph{arXiv
  preprint arXiv:2205.03519}, 2022.

\bibitem{Wang}
S.~Wang, Z.~Su, L.~Ying, X.~Peng, S.~Zhu, F.~Liang, D.~Feng, and D.~Liang,
  ``Accelerating magnetic resonance imaging via deep learning,'' in \emph{2016
  IEEE 13th International Symposium on Biomedical Imaging}.\hskip 1em plus
  0.5em minus 0.4em\relax IEEE, 2016, pp. 514--517.

\bibitem{liang2020deep}
D.~Liang, J.~Cheng, Z.~Ke, and L.~Ying, ``Deep magnetic resonance image
  reconstruction: Inverse problems meet neural networks,'' \emph{IEEE Signal
  Processing Magazine}, vol.~37, no.~1, pp. 141--151, 2020.

\bibitem{han2019k}
Y.~Han, L.~Sunwoo, and J.~C. Ye, ``k-space deep learning for accelerated mri,''
  \emph{IEEE Transactions on Medical Imaging}, vol.~39, no.~2, pp. 377--386,
  2019.

\bibitem{peng2022deepsense}
X.~Peng, B.~P. Sutton, F.~Lam, and Z.-P. Liang, ``Deepsense: Learning coil
  sensitivity functions for sense reconstruction using deep learning,''
  \emph{Magnetic Resonance in Medicine}, vol.~87, no.~4, pp. 1894--1902, 2022.

\bibitem{oh2020unpaired}
G.~Oh, B.~Sim, H.~Chung, L.~Sunwoo, and J.~C. Ye, ``Unpaired deep learning for
  accelerated mri using optimal transport driven cyclegan,'' \emph{IEEE
  Transactions on Computational Imaging}, vol.~6, pp. 1285--1296, 2020.

\bibitem{nakarmi2020multi}
U.~Nakarmi, J.~Y. Cheng, E.~P. Rios, M.~Mardani, J.~M. Pauly, L.~Ying, and
  S.~S. Vasanawala, ``Multi-scale unrolled deep learning framework for
  accelerated magnetic resonance imaging,'' in \emph{2020 IEEE 17th
  International Symposium on Biomedical Imaging (ISBI)}.\hskip 1em plus 0.5em
  minus 0.4em\relax IEEE, 2020, pp. 1056--1059.

\bibitem{huang2019deep}
P.~Huang, C.~Zhang, H.~Li, S.~K. Gaire, R.~Liu, X.~Zhang, X.~Li, and L.~Ying,
  ``Deep mri reconstruction without ground truth for training,'' in
  \emph{Proceedings of 27th Annual Meeting of ISMRM2019.}, 2019.

\bibitem{modl}
H.~K. Aggarwal, M.~P. Mani, and M.~Jacob, ``Modl: Model-based deep learning
  architecture for inverse problems,'' \emph{IEEE Transactions on Medical
  Imaging}, vol.~38, no.~2, pp. 394--405, 2018.

\bibitem{sun2016deep}
J.~Sun, H.~Li, Z.~Xu \emph{et~al.}, ``Deep admm-net for compressive sensing
  mri,'' \emph{Advances in Neural Information Processing Systems}, vol.~29,
  2016.

\bibitem{zhang2018ista}
J.~Zhang and B.~Ghanem, ``Ista-net: Interpretable optimization-inspired deep
  network for image compressive sensing,'' in \emph{Proceedings of the IEEE
  Conference on Computer Vision and Pattern Recognition}, 2018, pp. 1828--1837.

\bibitem{cui2021equilibrated}
Z.-X. Cui, J.~Cheng, Q.~Zhu, Y.~Liu, S.~Jia, K.~Zhao, Z.~Ke, W.~Huang, H.~Wang,
  Y.~Zhu \emph{et~al.}, ``Equilibrated zeroth-order unrolled deep networks for
  accelerated mri,'' \emph{arXiv preprint arXiv:2112.09891}, 2021.

\bibitem{dhariwal2021diffusion}
P.~Dhariwal and A.~Nichol, ``Diffusion models beat gans on image synthesis,''
  \emph{Advances in Neural Information Processing Systems}, vol.~34, pp.
  8780--8794, 2021.

\bibitem{nichol2021improved}
A.~Q. Nichol and P.~Dhariwal, ``Improved denoising diffusion probabilistic
  models,'' in \emph{International Conference on Machine Learning}.\hskip 1em
  plus 0.5em minus 0.4em\relax PMLR, 2021, pp. 8162--8171.

\bibitem{rombach2022high}
R.~Rombach, A.~Blattmann, D.~Lorenz, P.~Esser, and B.~Ommer, ``High-resolution
  image synthesis with latent diffusion models,'' in \emph{Proceedings of the
  IEEE/CVF Conference on Computer Vision and Pattern Recognition}, 2022, pp.
  10\,684--10\,695.

\bibitem{yang2022diffusion}
L.~Yang, Z.~Zhang, Y.~Song, S.~Hong, R.~Xu, Y.~Zhao, Y.~Shao, W.~Zhang, B.~Cui,
  and M.-H. Yang, ``Diffusion models: A comprehensive survey of methods and
  applications,'' \emph{arXiv preprint arXiv:2209.00796}, 2022.

\bibitem{DDPM}
J.~Ho, A.~Jain, and P.~Abbeel, ``Denoising diffusion probabilistic models,''
  \emph{Advances in Neural Information Processing Systems}, vol.~33, pp.
  6840--6851, 2020.

\bibitem{score-based}
Y.~Song and S.~Ermon, ``Generative modeling by estimating gradients of the data
  distribution,'' \emph{Advances in Neural Information Processing Systems},
  vol.~32, 2019.

\bibitem{score-based-SDE}
\BIBentryALTinterwordspacing
Y.~Song, J.~Sohl-Dickstein, D.~P. Kingma, A.~Kumar, S.~Ermon, and B.~Poole,
  ``Score-based generative modeling through stochastic differential
  equations,'' in \emph{International Conference on Learning Representations},
  2021. [Online]. Available: \url{https://openreview.net/forum?id=PxTIG12RRHS}
\BIBentrySTDinterwordspacing

\bibitem{jalal2021robust}
A.~Jalal, M.~Arvinte, G.~Daras, E.~Price, A.~G. Dimakis, and J.~Tamir, ``Robust
  compressed sensing mri with deep generative priors,'' \emph{Advances in
  Neural Information Processing Systems}, vol.~34, pp. 14\,938--14\,954, 2021.

\bibitem{song2022solving}
\BIBentryALTinterwordspacing
Y.~Song, L.~Shen, L.~Xing, and S.~Ermon, ``Solving inverse problems in medical
  imaging with score-based generative models,'' in \emph{International
  Conference on Learning Representations}, 2022. [Online]. Available:
  \url{https://openreview.net/forum?id=vaRCHVj0uGI}
\BIBentrySTDinterwordspacing

\bibitem{chung2022score}
H.~Chung and J.~C. Ye, ``Score-based diffusion models for accelerated mri,''
  \emph{Medical Image Analysis}, p. 102479, 2022.

\bibitem{kim2022diffusion}
B.~Kim and J.~C. Ye, ``Diffusion deformable model for 4d temporal medical image
  generation,'' in \emph{International Conference on Medical Image Computing
  and Computer-Assisted Intervention}.\hskip 1em plus 0.5em minus 0.4em\relax
  Springer, 2022, pp. 539--548.

\bibitem{cui2022self}
Z.-X. Cui, C.~Cao, S.~Liu, Q.~Zhu, J.~Cheng, H.~Wang, Y.~Zhu, and D.~Liang,
  ``Self-score: Self-supervised learning on score-based models for mri
  reconstruction,'' \emph{arXiv preprint arXiv:2209.00835}, 2022.

\bibitem{dar2022adaptive}
S.~U. Dar, {\c{S}}.~{\"O}zt{\"u}rk, Y.~Korkmaz, G.~Elmas, M.~{\"O}zbey,
  A.~G{\"u}ng{\"o}r, and T.~{\c{C}}ukur, ``Adaptive diffusion priors for
  accelerated mri reconstruction,'' \emph{arXiv preprint arXiv:2207.05876},
  2022.

\bibitem{nguyen2019frequency}
T.~Nguyen-Duc, T.~M. Quan, and W.-K. Jeong, ``Frequency-splitting dynamic mri
  reconstruction using multi-scale 3d convolutional sparse coding and automatic
  parameter selection,'' \emph{Medical Image Analysis}, vol.~53, pp. 179--196,
  2019.

\bibitem{blaimer2004smash}
M.~Blaimer, F.~Breuer, M.~Mueller, R.~M. Heidemann, M.~A. Griswold, and P.~M.
  Jakob, ``Smash, sense, pils, grappa: how to choose the optimal method,''
  \emph{Topics in Magnetic Resonance Imaging}, vol.~15, no.~4, pp. 223--236,
  2004.

\bibitem{10244070}
Y.~Guan, Y.~Li, R.~Liu, Z.~Meng, Y.~Li, L.~Ying, Y.~P. Du, and Z.-P. Liang,
  ``Subspace model-assisted deep learning for improved image reconstruction,''
  \emph{IEEE Transactions on Medical Imaging}, pp. 1--1, 2023.

\bibitem{kressner2014low}
D.~Kressner, M.~Steinlechner, and B.~Vandereycken, ``Low-rank tensor completion
  by riemannian optimization,'' \emph{BIT Numerical Mathematics}, vol.~54, pp.
  447--468, 2014.

\bibitem{9632354}
Z.~Ke, Z.-X. Cui, W.~Huang, J.~Cheng, S.~Jia, L.~Ying, Y.~Zhu, and D.~Liang,
  ``Deep manifold learning for dynamic mr imaging,'' \emph{IEEE Transactions on
  Computational Imaging}, vol.~7, pp. 1314--1327, 2021.

\bibitem{chung2022come}
H.~Chung, B.~Sim, and J.~C. Ye, ``Come-closer-diffuse-faster: Accelerating
  conditional diffusion models for inverse problems through stochastic
  contraction,'' in \emph{Proceedings of the IEEE/CVF Conference on Computer
  Vision and Pattern Recognition}, 2022, pp. 12\,413--12\,422.

\bibitem{tu2022wkgm}
Z.~Tu, D.~Liu, X.~Wang, C.~Jiang, M.~Zhang, Q.~Liu, and D.~Liang, ``Wkgm:
  Weight-k-space generative model for parallel imaging reconstruction,''
  \emph{arXiv preprint arXiv:2205.03883}, 2022.

\bibitem{peng2022one}
H.~Peng, C.~Jiang, Y.~Guan, J.~Cheng, M.~Zhang, D.~Liang, and Q.~Liu,
  ``One-shot generative prior learned from hankel-k-space for parallel imaging
  reconstruction,'' \emph{arXiv preprint arXiv:2208.07181}, 2022.

\bibitem{huang2023cdiffmr}
J.~Huang, A.~I. Aviles-Rivero, C.-B. Sch{\"o}nlieb, and G.~Yang, ``Cdiffmr: Can
  we replace the gaussian noise with k-space undersampling for fast mri?'' in
  \emph{International Conference on Medical Image Computing and
  Computer-Assisted Intervention}.\hskip 1em plus 0.5em minus 0.4em\relax
  Springer, 2023, pp. 3--12.

\bibitem{sarkka2019applied}
S.~S{\"a}rkk{\"a} and A.~Solin, \emph{Applied Stochastic Differential
  Equations}.\hskip 1em plus 0.5em minus 0.4em\relax Cambridge University
  Press, 2019, vol.~10.

\bibitem{anderson1982reverse}
B.~D. Anderson, ``Reverse-time diffusion equation models,'' \emph{Stochastic
  Processes and their Applications}, vol.~12, no.~3, pp. 313--326, 1982.

\bibitem{zbontar2018fastmri}
J.~Zbontar, F.~Knoll, A.~Sriram, T.~Murrell, Z.~Huang, M.~J. Muckley,
  A.~Defazio, R.~Stern, P.~Johnson, M.~Bruno \emph{et~al.}, ``fastmri: An open
  dataset and benchmarks for accelerated mri,'' \emph{arXiv preprint
  arXiv:1811.08839}, 2018.

\bibitem{knoll2020fastmri}
F.~Knoll, J.~Zbontar, A.~Sriram, M.~J. Muckley, M.~Bruno, A.~Defazio,
  M.~Parente, K.~J. Geras, J.~Katsnelson, H.~Chandarana \emph{et~al.},
  ``fastmri: A publicly available raw k-space and dicom dataset of knee images
  for accelerated mr image reconstruction using machine learning,''
  \emph{Radiology. Artificial Intelligence}, vol.~2, no.~1, 2020.

\bibitem{uecker2014espirit}
M.~Uecker, P.~Lai, M.~J. Murphy, P.~Virtue, M.~Elad, J.~M. Pauly, S.~S.
  Vasanawala, and M.~Lustig, ``Espirit—an eigenvalue approach to
  autocalibrating parallel mri: where sense meets grappa,'' \emph{Magnetic
  resonance in medicine}, vol.~71, no.~3, pp. 990--1001, 2014.

\bibitem{uecker2016bart}
M.~Uecker, J.~I. Tamir, F.~Ong, and M.~Lustig, ``The bart toolbox for
  computational magnetic resonance imaging,'' in \emph{Proc Intl Soc Magn Reson
  Med}, vol.~24, 2016, p.~1.

\bibitem{schlemper2017deep}
J.~Schlemper, J.~Caballero, J.~V. Hajnal, A.~N. Price, and D.~Rueckert, ``A
  deep cascade of convolutional neural networks for dynamic mr image
  reconstruction,'' \emph{IEEE Transactions on Medical Imaging}, vol.~37,
  no.~2, pp. 491--503, 2017.

\bibitem{sriram2020end}
A.~Sriram, J.~Zbontar, T.~Murrell, A.~Defazio, C.~L. Zitnick, N.~Yakubova,
  F.~Knoll, and P.~Johnson, ``End-to-end variational networks for accelerated
  mri reconstruction,'' in \emph{Medical Image Computing and Computer Assisted
  Intervention--MICCAI 2020: 23rd International Conference, Lima, Peru, October
  4--8, 2020, Proceedings, Part II 23}.\hskip 1em plus 0.5em minus 0.4em\relax
  Springer, 2020, pp. 64--73.

\bibitem{song2020improved}
Y.~Song and S.~Ermon, ``Improved techniques for training score-based generative
  models,'' \emph{Advances in Neural Information Processing Systems}, vol.~33,
  pp. 12\,438--12\,448, 2020.

\bibitem{SSIM}
Z.~Wang, A.~C. Bovik, H.~R. Sheikh, and E.~P. Simoncelli, ``Image quality
  assessment: from error visibility to structural similarity,'' \emph{IEEE
  Transactions on Image Processing}, vol.~13, no.~4, pp. 600--612, 2004.

\bibitem{levine2017fly}
E.~Levine and B.~Hargreaves, ``On-the-fly adaptive k-space sampling for linear
  mri reconstruction using moment-based spectral analysis,'' \emph{IEEE
  Transactions on Medical Imaging}, vol.~37, no.~2, pp. 557--567, 2017.

\bibitem{athalye2015parallel}
V.~Athalye, M.~Lustig, and M.~Uecker, ``Parallel magnetic resonance imaging as
  approximation in a reproducing kernel hilbert space,'' \emph{Inverse
  Problems}, vol.~31, no.~4, p. 045008, 2015.

\bibitem{haldar2016p}
J.~P. Haldar and J.~Zhuo, ``P-loraks: low-rank modeling of local k-space
  neighborhoods with parallel imaging data,'' \emph{Magnetic Resonance in
  Medicine}, vol.~75, no.~4, pp. 1499--1514, 2016.

\bibitem{nan2022data}
Y.~Nan, J.~Del~Ser, S.~Walsh, C.~Sch{\"o}nlieb, M.~Roberts, I.~Selby,
  K.~Howard, J.~Owen, J.~Neville, J.~Guiot \emph{et~al.}, ``Data harmonisation
  for information fusion in digital healthcare: A state-of-the-art systematic
  review, meta-analysis and future research directions,'' \emph{Information
  Fusion}, vol.~82, pp. 99--122, 2022.

\bibitem{NIPS2015_af473271}
C.~Chen, N.~Ding, and L.~Carin, ``On the convergence of stochastic gradient
  mcmc algorithms with high-order integrators,'' in \emph{Advances in Neural
  Information Processing Systems}, vol.~28, 2015.

\end{thebibliography}
\end{document}